\documentclass[sigconf,9pt]{acmart}

\newcommand{\PaperTitle}{Measuring Google AI Overviews:\\  Activation, Source Quality, Claim Fidelity, and Publisher Impact}

\usepackage{graphicx}
\usepackage{listings}
\usepackage{hyperref}
\usepackage{float}
\usepackage{url}
\usepackage{enumitem}
\usepackage{multirow}
\begin{document}
\acmYear{2026}\copyrightyear{2026}
\setcopyright{cc}
\setcctype[4.0]{by}
\acmConference[IMC '26]{Proceedings of the 2026 ACM Internet Measurement Conference}{October 12--16, 2026}{Karlsruhe, Germany}
\acmBooktitle{Proceedings of the 2026 ACM Internet Measurement Conference (IMC '26), October 12--16, 2026, Karlsruhe, Germany}
\acmDOI{10.1145/3777912.3839818}
\acmISBN{979-8-4007-2327-8/26/10}

\title{\PaperTitle}

\author{Haofei Xu}
\affiliation{%
  \institution{Washington University in St. Louis}
  \city{St. Louis}
  \state{MO}
  \country{USA}
}
\email{haofeix@wustl.edu}

\author{Umar Iqbal}
\affiliation{%
  \institution{Washington University in St. Louis}
  \city{St. Louis}
  \state{MO}
  \country{USA}
  }
\email{umar.iqbal@wustl.edu}

\author{Jacob M. Montgomery}
\affiliation{%
  \institution{Washington University in St. Louis}
  \city{St. Louis}
  \state{MO}
  \country{USA}
  }
\email{jacob.montgomery@wustl.edu}

\begin{abstract}
Google AI Overviews (AIOs) are arguably the most widely encountered deployment of generative AI, reaching over 2 billion users who may not realize the answers they see are AI-generated.
Where search engines have traditionally surfaced ranked sources and left users to evaluate them, AIOs synthesize and deliver a single answer --- giving Google unprecedented editorial control over what users read and know.
We present a large-scale longitudinal measurement study, issuing 55,393 trending queries across 19 topical categories over a 40-day window (March~13--April~21, 2026).
We report four main findings.
First, overall AIO activation is 13.7\%, rising to 64.7\% for question-form queries, while politically sensitive topics see markedly lower rates.
Second, AIO-cited domains are more credible than co-displayed first-page results, yet nearly 30\% do not appear in those results at all, indicating a source selection mechanism distinct from Google's ranking algorithm.
Third, decomposing responses into 98,020 atomic claims, 11.0\% are unsupported by the cited pages --- with omission the dominant failure mode --- and source quality and claim fidelity are largely independent.
Fourth, well over half of AIO-cited pages carry display advertising, meaning publishers lose revenue when AIOs suppress the click-through, even as Google's own sponsored ads continue to appear on the same page.
Together, these findings document a rapid transformation of the online information ecosystem whose consequences for epistemic security remain poorly understood.
\end{abstract}



\maketitle

\section{Introduction}

Every day, billions of people turn to Google Search to answer their most consequential questions --- about their health, their finances, the news, and the world.
For decades, search meant a ranked list of sources: Google chose what to surface, but users chose what to read and whom to trust.
That paradigm is now fundamentally changing.
Google's AI Overviews (AIOs) synthesize information drawn from across the web and deliver a single, prose answer positioned above every other result on the page.
The user sees a summary, not a list of sources to choose from.

Google reports that more than 2 billion users now encounter AIOs worldwide~\cite{pichai2025alphabetq2earnings}, a figure that has grown since that estimate was published in July 2025.
Unlike a chatbot  --- interfaces that users actively choose --- AIOs are injected into the default search experience.
Many users encountering them may not realize the answer they are reading is machine-generated.
This makes AIOs arguably the most consequential deployment of generative AI in existence: not necessarily the most capable, but by far the most widely and passively encountered.
The epistemic stakes are correspondingly large.
When a single system authors the answer that hundreds of millions of people read in response to a health query, a legal question, or a political search, the appearance, accuracy, and sourcing of that answer matter enormously --- and so does the question of who profits.

Google describes AIOs as ``built to surface information that is backed up by top web results'' and as generally free of the hallucinations that affect other LLM systems~\cite{google_whitepaper_2025} --- empirically testable claims that independent research has not yet fully evaluated.
Existing studies are either narrow in domain~\cite{hu2026auditing}, limited to behavioral outcomes~\cite{agarwal2026field,khosravi2026impact,xu2025aisummaries}, or based on fixed query sets collected before AIOs reached their current scale and deployment maturity~\cite{aral2026rise}.
Indeed, no prior work simultaneously characterizes AIO activation patterns, source selection, claim fidelity, and publisher economic exposure across a diverse, naturalistic query sample over a sustained period.

We present a large-scale longitudinal measurement study to fill this gap.
Over 40 days (March~13--April~21, 2026), we issued 55,393 trending queries spanning 19 topical categories using a distributed Puppeteer-based crawler deployed on AWS Lambda, capturing each AIO, its embedded reference citations, the co-displayed first-page results, and the full content and advertising infrastructure of every cited webpage.
We characterize AIO behavior across four dimensions that span the system's operation from trigger to consequence.

We report four main findings.
First, overall AIO activation is 13.7\%, but this aggregate conceals sharp structural variation: question-form queries trigger AIOs at 64.7\% versus 9.5\% for non-question queries, a 6.8 times difference, while politically sensitive categories see markedly suppressed rates, suggesting undisclosed editorial discretion in Google's triggering logic.
Second, contrary to prior work showing AIOs draw on lower-quality sources~\cite{aral2026rise}, we find AIO-cited domains are systematically more credible than co-displayed first-page results; however, 29.8\% of AIO-cited domains do not appear in those first-page results at all, indicating that Google's AIO system draws on a source pool and prioritization system distinct from its own ranking algorithm.
Third, decomposing AIO responses into 98,020 atomic claims and verifying each against its cited sources, 4.1\% are directly contradicted by or conflict with their cited content, while a further 7.0\% are not addressed in the source text our pipeline could retrieve.
Fourth, more than half (at least 50.6\%) of AIO-cited pages carry display advertising while Google's own sponsored search ads continue to appear on AIO-bearing pages --- in some cases above the AIO itself --- suggesting that AIO deployment displaces publisher revenue while preserving Google's own ad capture.

Taken together, these findings provide a systematic empirical characterization of a widely deployed AI-mediated information system whose internal operation has been largely opaque.
Several of our findings stand in direct tension with Google's own public claims, and suggest that continued independent measurement will be necessary to understand how the system operates and what its effects are on the broader information ecosystem.

To foster follow-up research, we release our code and data at
https://github.com/haofeix/google-aio-characterization
\section{Background and Motivation}
\label{sec:background}

\subsection{Background}
\label{sec:background:bg}

\paragraph{What AIOs are and how they are generated.}
Google AI Overviews are not standalone LLM responses.
Rather, they are powered by a custom Gemini model that operates in tandem with Google's traditional ranking infrastructure and Knowledge Graph~\cite{google_whitepaper_2025}.
AI Overviews are therefore better understood as search-integrated generative summaries that synthesize responses from supporting web content, aligning them more closely with retrieval-augmented generation than with chatbot responses generated from parametric memory alone~\cite{google_whitepaper_2025}.
Each AIO is accompanied by embedded reference citations that are presented to the user as the evidentiary basis for the generated content; these citations are structural rather than incidental, because the design intent is that every claim should be traceable to a specific retrieved source.
This architecture is central to Google's quality argument. Because the model is constrained to synthesize from retrieved pages, Google positions AIOs as meaningfully less prone to the fabrication errors that have plagued general-purpose LLMs~\cite{google_whitepaper_2025}.
Whether the grounding architecture delivers on that promise is one of the central empirical questions this paper addresses.

\paragraph{How triggering works.}
AIOs do not appear on every query.
Google's systems determine when generative AI ``adds additional benefit beyond what people might already get on Search today'' and where it has ``high confidence in the quality of the responses''~\cite{google_whitepaper_2025}.
In practice this means AIOs are more prevalent for informational and question-form queries, and are suppressed for queries involving ``sensitive, explicit, rapidly evolving, or dangerous topics''~\cite{google_whitepaper_2025}.
Google acknowledges this suppression in broad strokes but does not publish the criteria, thresholds, or topical categories that govern when an AIO appears or is withheld.
The triggering logic is entirely opaque to outside observers, which means the only way to characterize it is empirically, at scale.

\paragraph{Source selection and SERP positioning.}
Each AIO contains embedded reference links that are presented to the user as the basis for the generated content.
Google describes these citations as drawn from ``top web results,''~\cite{google_whitepaper_2025} language that implies the AIO's source pool is closely related to or drawn from the set of pages that Google's ranking algorithm surfaces.
Whether this is true in practice --- and what fraction of AIO-cited pages do not appear in the co-displayed first-page results for the same query --- is not publicly documented.
Beyond source selection, AIOs occupy a privileged position on the search engine results page (SERP): they are rendered above all other results and, in some configurations, above or alongside Google's own sponsored search advertisements.
This positioning has direct consequences for both what users read first and what pages they interact with.

\paragraph{Scale, trajectory, and the publisher ecosystem.}
AIO deployment has expanded rapidly since launch, growing from seven countries in 2024 to 229 countries by 2025, with trigger rates rising  commensurately~\cite{aral2026rise}.
This expansion has coincided with measurable reductions in publisher traffic.
A randomized field experiment finds that AIOs reduce organic click-throughs by 38\% and increase zero-click searches by 33\%, with no compensating improvement in user-reported satisfaction~\cite{agarwal2026field}.
Complementary quasi-experimental evidence finds that AIO exposure reduces traffic to English Wikipedia articles by approximately 15\%~\cite{khosravi2026impact}.
These behavioral shifts have real economic consequences for publishers who depend on search referral traffic~\cite{geo_kdd}.
A Digital Content Next survey of premium publishers found a median year-over-year decline in Google Search referral traffic of 10\% over just eight weeks in mid-2025. 
The number of publishers reporting declines was twice the number reporting gains~\cite{dcn2025}.
Individual cases are more severe: music publisher Stereogum attributed the loss of 70\% of its advertising revenue to AIOs~\cite{adexchanger2026}, and travel blog The Planet D shut down entirely after its traffic fell 90\% following AIO rollout~\cite{npr2025}.
Because advertising is the dominant revenue model for digital publishers, traffic displacement maps directly onto revenue displacement~\cite{pew2025_ai_clicks} --- a concern made more acute by the fact that Google's own advertising revenue has continued to grow over the same period~\cite{alphabet10k2025}.

These tensions have drawn regulatory scrutiny across multiple jurisdictions, with the European Commission~\cite{ec2025} and the UK CMA~\cite{cma2026} targeting AI Overviews directly, and U.S. courts acknowledging publisher harms in ongoing antitrust proceedings~\cite{doj2025}.
Independent empirical characterization of AIO behavior is therefore not merely an academic exercise --- it bears directly on the evidentiary questions regulators are asking.
\subsection{Motivating Concerns and Research Questions}
\label{sec:background:rqs}

\paragraph{RQ1 --- Activation.}
The most fundamental question about any deployed information system is: who encounters it, and under what conditions?
For AIOs, this means understanding the triggering logic --- which queries activate an AIO and which do not.
This matters for two reasons.
First, activation determines exposure. A user who poses a natural question is far more likely to receive an AI-generated answer than one who types a short keyword string, which suggests that less search-experienced users may be more exposed to AI-generated content than more experienced ones~\cite{vonhoyer2022search,bilal2018children}.
Second, selective suppression of AIOs on certain topic categories raises questions about undisclosed editorial discretion that are not publicly acknowledged by Google.
The company states that AIOs are suppressed for sensitive, explicit, or dangerous topics, but the boundaries of that suppression --- and whether they are applied consistently across topics --- are undocumented.
Our first research question is therefore: \emph{how does AIO activation vary across query structure, topical category, and query length, and what does that variation reveal about Google's undisclosed triggering logic?}

\paragraph{RQ2 --- Source quality.}
Google's central quality claim is that AIOs draw on high-quality, reliable sources grounded in top web results.
Prior work has tested this claim using categorical credibility taxonomies.
Using the Media Bias/Fact Check database, Aral et al.~\cite{aral2026rise} found that sources cited by AIOs are, on average, less credible than those returned by traditional search.
This finding stands in direct tension with Google's own characterization. 
Categorical credibility measures, however, assign sources to broad classes and do not distinguish among sources within the same class.
For example, a news organization with rigorous fact-checking and a partisan news outlet may both be labeled ``editorially reviewed,'' even though their reliability differs substantially.
By contrast, continuous credibility measures assign each domain a numerical score, capturing these differences and enabling more precise comparisons between the credibility distributions of AIO references and first-page search results.
We therefore revisit this finding using a diverse query sample and PC1~\cite{lin2023high}, a continuous measure that provides a finer-grained assessment of source quality.

A second and independent concern is whether AIO citations overlap with first-page SERP results at all: if a substantial fraction of AIO-cited domains do not appear in the SERP results for the same query, then Google's AIO system is drawing from a source pool that its own ranking algorithm does not surface --- a transparency concern that is independent of whether those sources are credible.
Our second research question is: \emph{how does the credibility of AIO-cited domains compare to co-displayed first-page results, and to what extent do these two source pools overlap?}

\paragraph{RQ3 --- Claim fidelity.}
Google explicitly states that AIOs ``generally don't hallucinate in the ways that other LLM experiences might''~\cite{google_whitepaper_2025}, attributing this to the grounding architecture described above (Section~\ref{sec:background:bg}).
However, prior audits of other AI search systems have found high unsupported-claim rates \cite{liu_emnlp}: a 2025 evaluation of multiple LLMs with web access found that between 50\% and 90\% of response statements were not fully supported by the cited sources~\cite{wu_2025, aral2026rise}, and an audit of ChatGPT Search found confident misattribution in the majority of tested cases~\cite{jazwinska2024chatgpt}.
Whether AIOs, with their tighter grounding architecture, perform substantially better is an open empirical question.
Equally important is the relationship between source quality and claim fidelity.
If high-quality sources reliably produce grounded claims and low-quality sources do not, then improving sourcing would address the fidelity problem.
If source quality and claim fidelity are largely independent, however, the problem is structural and cannot be fixed at the sourcing layer alone.
Our third research question is: \emph{what fraction of atomic claims in AIO responses are unsupported by their cited sources, what forms do these inconsistencies take, and is claim fidelity correlated with source quality?}

\paragraph{RQ4 --- Publisher economics.}
The economic concern around AIOs is now well documented at the aggregate level.
Causal evidence from Agarwal \& Sen~\cite{agarwal2026field} establishes that AIOs reduce organic click-throughs by 38\%, while Khosravi \& Yoganarasimhan~\cite{khosravi2026impact} find a 15\% decline in Wikipedia traffic attributable to AIO exposure.
What has not been characterized is the economic structure on the publisher side of this displacement: specifically, how many AIO-cited pages carry display advertising and whether a suppressed click therefore represents a meaningful and quantifiable revenue loss.
Equally important is the asymmetry on Google's side.
If Google's own sponsored search advertisements continue to appear on AIO-bearing pages --- and in some cases appear above the AIO itself --- then AIO deployment is asymmetric in its economic consequences: it suppresses the organic clicks that drive publisher ad revenue while leaving Google's own ad inventory intact.
Our fourth research question is therefore: \emph{what is the advertising exposure of AIO-cited publisher pages, and how does AIO placement interact with Google's own sponsored search inventory on the same SERP?}

\subsection{Related Work}
A substantial literature establishes that online platforms function as epistemic infrastructure \cite{ge2025does}, shaping what users believe through ranking, aggregation, and content selection~\cite{epstein2015search,robertson2018auditing,vosoughi2018spread}, and work on retrieval-augmented generation shows that grounding LLM outputs in web content does not eliminate hallucination or citation errors~\cite{venkit2024searchenginesaiera,magesh2025hallucinations}.
A much smaller body of work examines AIOs directly.
Domain-specific audits document quality concerns in health and baby care queries, including inconsistencies between AIO responses and cited sources~\cite{hu2026auditing}.
Aral et al.~\cite{aral2026rise} measured AIO activation across 243 countries but relied on fixed benchmark queries and discrete snapshots collected before AIOs reached their current scale.
Two recent studies document the behavioral consequences of AIO deployment --- including a 38\% reduction in organic clicks and a 33\% increase in zero-click searches~\cite{agarwal2026field,khosravi2026impact}.
Another controlled experiment with mock search pages further shows that exposure to AI summaries shifts users' attitudes and policy support~\cite{xu2025aisummaries}. 
But none of these works characterizes the underlying system properties that produce those effects.
Ours is the first study to simultaneously characterize AIO activation, source selection, claim fidelity, and publisher economic exposure at scale across a diverse, naturalistic query sample.
\section{Methodology}
\label{sec:methodology}

\begin{figure*}[t]
    \centering
    \includegraphics[width=\textwidth]{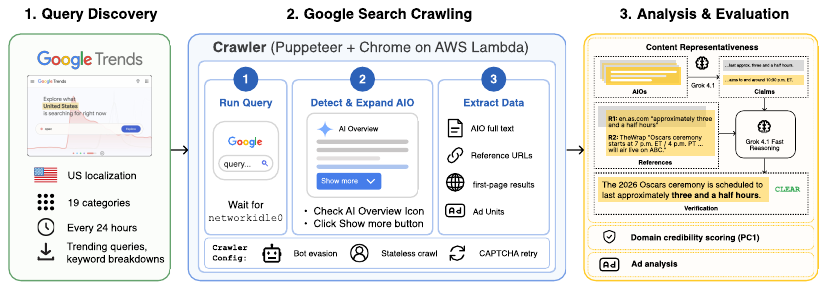}
    \caption{Our approach to characterizing the Google AIO ecosystem: (1) We begin by extracting top search queries from Google Trends. (2) We develop a bespoke crawler to capture search results for these queries, including their AIOs and associated metadata (e.g., reference links). We then crawl the referenced pages to extract their content and detect the presence of advertisements. (3) We evaluate the quality of AIOs by assessing the credibility of cited references and the consistency of AIO claims with source content. Finally, we analyze the potential economic impact of AIOs on the web ecosystem.}
    \label{fig:pipeline}
\end{figure*}

This section describes our methodology for characterizing the Google 
AIO ecosystem.
We identify trending Google search queries, issue them at scale, and 
collect the resulting AIO content alongside associated metadata --- 
including cited links and the content of the corresponding webpages.
We then analyze key characteristics of the results and examine the 
potential economic impact of AIOs on the broader web ecosystem.
Figure~\ref{fig:pipeline} provides an overview of our approach.

\subsection{Web Crawler}

We first need a web crawler that allows us to extract the top-searches on Google, trigger and capture their AIOs and associated metadata, as well as scrape the content of the references mentioned in AIOs.
To that end, we design a Puppeteer-based web crawler~\cite{puppeteer}. 
Our crawler uses Google Chrome as the underlying browser, and it is deployed as a set of AWS Lambda functions in the us-east-1 AWS region, located in Northern Virginia, U.S. 
As standard search results vary by geolocation~\cite{hannak2013measuring}, AIO activation, content, and cited sources may similarly exhibit geographic variation.
Since our measurements originate from a single U.S.-based geographic and network vantage point, we scope our findings to the U.S.-localized AIO experience and leave cross-region comparisons to future work.

Each crawl instance spawns a stateless, fresh browser profile (i.e., without cookies, cache, and session data carry over between requests) to ensure that prior browsing history does not influence the search engine results pages we observe.
To reduce the likelihood of bot detection, we spoof relevant browser properties (i.e., remove the \texttt{webdriver} navigator flag, present a standard macOS Chrome user-agent), type search queries character-by-character with randomized inter-keystroke delays, and introduce variable pauses between page interactions.
We do not attempt to bypass CAPTCHAs.
When a CAPTCHA wall is encountered, we re-queue the request for a fresh crawl, up to a maximum of 20~retries.

\subsection{Collecting Google AIOs and Metadata}

\subsubsection{Google Search Results}
We use Google's publicly available trending search queries as the basis of our dataset, as they capture queries currently popular or experiencing significant increases in user interest, providing a useful proxy for real-world search behavior.
We issue these queries to Google and collect the corresponding AIOs and associated metadata, as described below.

\paragraph{Search Trends}
We use our crawler to navigate to the US-localized Google Trends dashboard and iterate across 19~predefined topical categories\footnote{The categories are: Autos \& Vehicles, Beauty \& Fashion, Business \& Finance, Climate, Entertainment, Food \& Drink, Games, Health, Hobbies \& Leisure, Jobs \& Education, Law \& Government, Other, Pets \& Animals, Politics, Science, Shopping, Sports, Technology, and Travel \& Transport.}, every 24~hours.
For each category, we export the trending queries along with their search volumes and keyword breakdowns.

\paragraph{AI Overviews}
We next run each search term in our dataset on Google search webpage in a fresh, stateless browser instance. 
Our crawler waits for the \texttt{networkidle0}~\cite{puppeteer-waituntil} event to trigger (i.e., no outstanding network requests for 500 ms), then proceeds to identify whether an AIO is present on the rendered search results webpage.
As Google dynamically and periodically changes the HTML structure of the results page, including HTML tag names and attribute values, it is non-trivial to extract AIO content from a fixed selector.
We therefore adopt a rule-based anchoring strategy.

First, we locate the AIO icon, which is an \texttt{<svg>} element (24$\times$24\,px) whose nearest parent \texttt{<div>} contains a direct child with \texttt{textContent} containing ``AI Overview''.
When an AIO's presence is confirmed, we click the ``Show more'' button to fully expand the content, since Google's search results webpage uses lazy loading to truncate long summaries by default. 
We then extract the full text from the \texttt{<div>} elements that contain the AIO icon.
For seven queries, the container included the message ``An AI Overview is not available for this search'' within a hidden HTML element that was not visible to the user.
We exclude these queries from our analysis.
For each AIO, we also collect the full set of embedded reference citations at the AIO level, recording titles, URLs, and cited text fragments (parsed from \texttt{span[data-huuid]} elements).
In addition, we capture all first-page SERP URLs.

\subsubsection{Webpage Content}
\label{sec:webpage_content}
To verify claim consistency, we scrape the textual content of all webpages cited in AIO references.
Each page is parsed using \texttt{readability-lxml}~\cite{readability-lxml} to isolate the primary article body, removing boilerplate elements such as headers, footers, and navigation menus, then converted to plain text using \texttt{BeautifulSoup4}~\cite{beautifulsoup4}.
Pages that fail this initial pass are re-fetched using a headful Chrome instance with an authenticated browser profile to reduce bot detection.
PDF references are downloaded and converted to plain text using \texttt{pdftotext} from Poppler~\cite{poppler}.

We exclude references hosted on short-form social media platforms (i.e., Facebook, Instagram, LinkedIn, Pinterest, Reddit, Threads, TikTok, X/Twitter, and YouTube.), as these services often require authentication and employ bot-detection mechanisms that are difficult to reliably bypass in an automated pipeline.

A small number of pages (n=262; $<$1\%) contain paywalls.
In such cases, we retain only the visible portion --- typically the headline and leading paragraphs --- since AIO-referenced claims are often introduced in the lede. We failed to capture text for less than 1\% of pages we attempted to collect.
Nonetheless, claims relying on content behind the paywall may be marked unsupported by our consistency checker (Section~\ref{sec:methodology-representativeness}).

\subsubsection{Ad Extraction}
To measure the economic exposure of AIO-cited publishers, we parse the DOM of each referenced webpage and count \texttt{<iframe>} elements matching EasyList filter rules~\cite{easylist}, a standard approach used extensively in prior work~\cite{yeung2023onlineads,zeng2020bad,iqbal2020adgraph}. We also capture the full SERP whenever an AIO is present and extract sponsored ads by identifying elements labeled ``Sponsored,'' classifying each as appearing above or below the AIO.

\subsection{Checking Content Integrity}
\label{sec:methodology-representativeness}
We analyze AIO content quality along two dimensions: the credibility of cited reference URLs, and the consistency between claims made in AIOs and the content of those references.
We describe each in turn.

\subsubsection{Reference URL Quality Scoring}
\label{sec:doamin-quality}

We assess source credibility using the continuous PC1 score from Lin et al.~\cite{lin2023high}, derived as the first principal component of multiple expert and crowd ratings of news-domain credibility.  This results in a continuous metric ranging from 0 to 1 (e.g., \texttt{reuters.com}~$= 1.000$, \texttt{nytimes.com}~$= 0.859$, 
\texttt{facebook.com}~$= 0.407$), covering 11{,}520 domains in total.

We resolve each reference URL to a score via a three-level lookup: (i)~longest-prefix \emph{path} match, which distinguishes path-level entries the index defines separately from their root (e.g., \texttt{facebook.com/news}~$=0.833$ vs.\ \texttt{facebook.com}~$=0.407$); (ii)~exact host match (after stripping leading \texttt{www.}); and (iii)~subdomain fallback, peeling one leading subdomain at a time down to two remaining segments, which avoids collapsing into ccTLDs like \texttt{co.uk} while recovering, e.g., \texttt{en.wikipedia.org}~$\to$~\texttt{wikipedia.org}.
\subsubsection{Checking Reference Consistency}
Source credibility alone does not guarantee AIO accuracy, given the well-known limitations and failure modes of generative AI (e.g., hallucinations)~\cite{tong-etal-2024-llms}.
We therefore adopt a conservative criterion for quality: we evaluate whether each AIO faithfully represents the content of its cited sources, decomposing AIO responses into atomic claims and verifying each against the corresponding reference texts. 
\paragraph{LLM-based Consistency Checking Framework}

We develop an LLM-based framework to check claim consistency, building on prior work~\cite{min2023factscore, wu2025gpts, cite_eval}.
Our approach decomposes the problem into two sequential subtasks: atomic claim extraction, in which AIO responses are broken into individual factual assertions, and claim verification, in which each assertion is checked against the content of its cited source.
We curate separate few-shot prompts~\cite{brown2020language} for each subtask, as task-specific examples have been shown to substantially improve LLM performance~\cite{brown2020language,gao2021making,wei2022chain}.
We use Grok~4.1~Fast~Reasoning~\cite{xai-grok-4-1} as the underlying model with \texttt{temperature} set to 0 to reduce stochastic variation when applying the same fixed prompts across the dataset.
We evaluate the framework’s reliability through human validation of both pipeline stages (Section~\ref{sec:human-annotation}).

\paragraph{Claim Extraction}
\label{sec:claim-extraction}
We first decompose each AIO into atomic, self-contained factual claims.
Our approach is to split compound sentences into one-fact-per-claim units, rewrite each claim to be self-contained (replacing pronouns with the full entity names they refer to elsewhere in the AIO), drop section headings, and de-duplicate repeating facts, as also explored by prior work~\cite{metropolitansky2025claimify}.
The full prompt and the translation of an example AIO to its atomic claims are in Appendix~\ref{appendix:claim-extraction-prompt}.

\begin{table*}[!t]
\footnotesize
\centering
\begin{tabular}{lp{6.6cm} p{8cm}}
\toprule
\textbf{Label} & \textbf{AIO Claim} & \textbf{Reference Evidence} \\
\midrule
\textsc{Clear} &
Ryan Coogler won his first career Oscar for Best Original Screenplay for \emph{Sinners} at the 98th Academy Awards. & ``Ryan Coogler won his first career Oscar\ldots for best original screenplay award for `Sinners' '' (direct confirmation from a source).\\ \midrule
\textsc{Vague} &
The 2026 Oscars are scheduled to end around 10:30 p.m.\ ET on Sunday, March 15, 2026. & Date and time clearly stated. No direct mention of end time. One source mentions ``ending around 10:30 p.m.\ ET'' but for the 2025 edition, another source mentions time generally averaging ``three and a half hours''. \\ \midrule
\textsc{Ambiguous} &
\emph{Good Grief} is Dan Levy's 2024 feature directorial debut. & ``...in 2023 with \emph{Good Grief}...''---\textit{The Hollywood Reporter}, ``...2024's \emph{Good Grief}...''---\textit{Netflix} (sources contradicting each other on the release year) \\ \midrule
\textsc{Incorrect} &
Artemis~II is a crewed mission that orbits the Moon. & ``[it] will not land on the moon or even go into lunar orbit. Instead, the plan is to loop around it'' (references describe a \emph{flyby}, not an orbit) \\ \midrule
\textsc{Omitted} &
Tickets for Coachella 2026 are available on AXS. &
No cited reference mentions AXS. References name the official Coachella site, StubHub, etc. \\
\bottomrule
\end{tabular}
\caption{Examples of each enumerated claim consistency type. \textit{AIO Claim} shows claim statements from AIOs and \textit{Reference Evidence} shows the specific statement in the referenced content that support or oppose the claim.}
\label{label_examples}
\end{table*}

\paragraph{Claim Verification}
\label{sec:claim_verification}
Once we have extracted the claims, we check whether the extracted claims are supported by the reference content. 
Rather than treating support as a binary condition, we recognize that consistency exists along a spectrum.
This problem parallels consistency analysis in privacy policies where the goal is to compare a specific claim against a larger reference document and characterize the type and degree of support it receives. 
Prior work on privacy policy consistency analysis has addressed this challenge through more nuanced labeling schemes.
Building on prior work~\cite{polisis,PoliCheck,trimananda2022ovrseen,PoliGraph,wu2025gpts}, we assign each claim to one of the following categories:
\textsc{Clear:} the claim is directly and explicitly supported by the reference content.
\textsc{Vague:} the claim is supported, but only in broader or less specific terms.
\textsc{Ambiguous:} the claim receives conflicting support from different parts of the reference content.
\textsc{Incorrect:} the claim is contradicted by the reference content.
\textsc{Omitted:} the claim is not supported by the reference content.
We further group these labels into two high-level categories: \textit{consistent} (\textsc{Clear} and \textsc{Vague}) and \textit{inconsistent} (\textsc{Ambiguous}, \textsc{Incorrect}, and \textsc{Omitted}).
Table~\ref{label_examples} illustrates each label with examples drawn from our dataset.

\begin{table}[t]
\centering
\small
\begin{tabular}{lrr}
\toprule
\textbf{Label} & \textbf{Count} & \textbf{Percentage} \\
\midrule
\textbf{C}lear     & 82,933 & 84.6\% \\
\textbf{V}ague     &  4,271 &  4.4\% \\
\textbf{A}mbiguous &  1,367 &  1.4\% \\
\textbf{I}ncorrect &  2,609 &  2.7\% \\
\textbf{O}mitted   &  6,840 &  7.0\% \\
\midrule
\textbf{Consistent (C + V)}   & \textbf{87,204} & \textbf{89.0\%} \\
\textbf{Inconsistent (A + I + O)}     & \textbf{10,816} & \textbf{11.0\%} \\
\bottomrule
\end{tabular}
\caption{Overall verification label distribution across 98,020 claims from 7,491 verifiable AIOs out of 7,583 total AIOs (92 omitted: 80 with zero extractable claims + 12 with only social-media references).}
\label{tab:verification_label_distribution}
\end{table}

To condition the LLM to assign claim consistency labels, we provide several examples of \texttt{<claim description,} \texttt{relevant reference content statement,} \texttt{consistency label>} tuples in our prompt. 
For each AIO, we pass a numbered list of extracted claims, relevant text that is embedded in the reference URL as a text fragment (i.e., \texttt{\#:$\sim$:text=}), and the full extracted body text of every cited reference, each labeled with an identifier (\texttt{[R1]}, \texttt{[R2]}, \dots, \texttt{[RN]}) and its source URL. 
Note that we verify claims against the full extracted reference text, the text fragments are provided as additional context to help identify the passage Google intended to cite.

For the output, we instruct our framework to provide a confidence score, the matched references, and a supporting evidence span for each claim. 
We provide the full prompt in Appendix~\ref{app:verify_claims_prompt}. 

\subsubsection{Pipeline Validation}
\label{sec:human-annotation}

We validate both stages of the pipeline --- claim extraction and claim verification --- against independent human annotation.

\paragraph{Claim extraction.}
Two annotators (graduate students, including the first author) independently labeled every extracted claim from 100 randomly sampled AIOs, applying four codebook rules: \textit{W1 (faithfulness)}, the claim must be supported by the visible AIO text; \textit{W2 (atomicity)}, the claim must state a single proposition; \textit{W3 (self-containment)}, the claim must be interpretable on its own without unresolved references; and \textit{W4 (no meta-claims)}, the claim must assert a fact about the world rather than about the source itself.
Annotators also recorded any atomic facts in the AIO not covered by any extracted claim.
Inter-annotator agreement was high (Cohen's $\kappa = 0.85$ on per-claim wrong flags; $\mathrm{ICC}(2,1) = 0.92$ for per-sample wrong counts), with remaining disagreements resolved by adjudication.
Across 1,379 sampled claims, only 24 (1.74\%) were wrong (W1: 0, W2: 4, W3: 12, W4: 8), and 273 atomic facts were missing, yielding precision 98.26\%, recall 83.23\%, and $F_1$ 90.12\%.
Omission rather than fabrication is the extractor's dominant error mode: 62 of 100 samples contained a missing fact, while only 15 contained a wrong claim.

\paragraph{Claim verification.}
We validated the verifier on a stratified sample of 100 claim-level verdicts (20 per label), ensuring each of the five categories was audited at adequate scale despite their unequal base rates.
Two annotators independently re-labelled each verdict against the same matched references using the five-class decision rules: \textsc{Clear} requires literal support; \textsc{Vague} permits inferred but not literal support; \textsc{Ambiguous} requires two cited references to conflict; \textsc{Incorrect} requires direct contradiction by a cited reference; and \textsc{Omitted} applies when no cited reference addresses the claim's substance.
Inter-annotator agreement was almost perfect (Cohen's $\kappa = 0.94$), with remaining disagreements resolved by adjudication.
After adjudication, the verifier matched the human label on 98 of 100 verdicts, with per-class accuracy of 95\% for \textsc{Clear} and \textsc{Vague}, and 100\% for \textsc{Ambiguous}, \textsc{Incorrect}, and \textsc{Omitted}.
Both errors involved the verifier being too generous with supported claims, not too harsh --- it never wrongly flagged a grounded claim as \textsc{Incorrect} or \textsc{Omitted}, and human reviewers agreed perfectly on all three failure-mode labels that drive our fidelity results (\textsc{Ambiguous}, \textsc{Incorrect}, and \textsc{Omitted}).
Weighted by empirical label frequencies in Table~\ref{tab:verification_label_distribution}, overall verifier accuracy is 95.6\%.
Although the high inter-annotator agreement and adjudication procedure reduce sensitivity to individual judgments, they do not eliminate possible sampling or evaluator bias.

\section{Evaluation}
\label{sec:eval}

We organize our results around the four dimensions introduced in Section~\ref{sec:background}: AIO activation patterns, source selection, claim fidelity, and publisher economic exposure.

\subsection{AIO Activation}

\begin{figure}[t]
\centering
\includegraphics[width=\columnwidth]{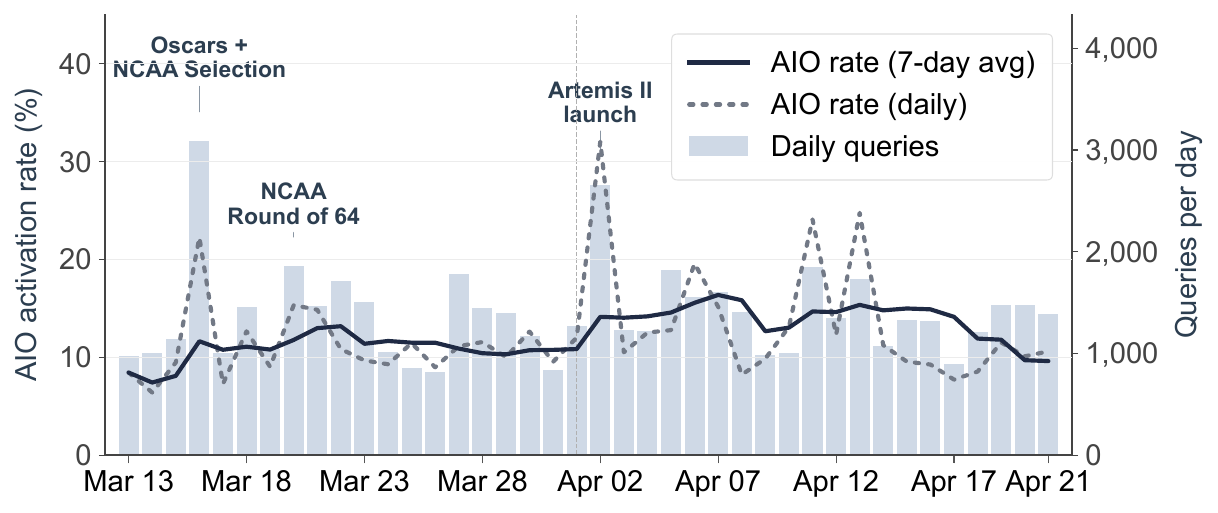}
\caption{Daily AIO activation rate (solid line: trailing 7-day moving average; dotted line: raw daily rate) and daily query volume (light blue bars, right axis) over the 40-day observation window (Mar 13--Apr 21, 2026; 55{,}393 queries; 7{,}583 AIOs). Annotations mark the three days with the highest query volume.}
\label{fig:daily_activation}
\end{figure}

Across the 40-day observation window (March~13--April~21, 2026), our crawler issued 55,393 trending queries spanning all 19 topical categories, of which 7,583 (13.7\%) returned an AIO.
The day-to-day activation rate is broadly stable with no monotonic trend, but superimposed on this baseline are sharp, short-lived spikes coinciding with major public events.

The three highest-volume days were March~16 (Oscars and NCAA tournament selection), April~2 (Artemis II launch), and March~20 (NCAA Round of 64), with query volumes approximately 2.2, 1.9, and 1.3 times the daily mean, respectively (Figure~\ref{fig:daily_activation}). 
While these dates were identified on the basis of query volume, March~16 and April~2 also exhibited elevated AIO activation rates (22.3\% and 32.1\%, respectively). 
However, this co-occurrence should not be interpreted as a general volume effect across categories: across the 19 topical categories, query count was not significantly correlated with category-specific AIO activation rate (Spearman's $\rho=0.02$, $p=0.93$).
Although such concentration may be relatively low-stakes for sports and entertainment events, it could be more consequential during high-stakes public events, such as elections or public-health emergencies, when users are seeking timely and rapidly evolving information.

We also observe that some queries repeat in our dataset; for example, a news event may have the same query trending over several days.
To examine how query repetition influences AIO activation rate, we de-duplicate such queries, and note that the activation rate changed from 13.69\% to 16.03\% (details in Appendix~\ref{app:repeated_queries}). 

\paragraph{Topical variation.}
AIO activation varies by a factor of 13 across categories, from 3.5\% in 
Beauty \& Fashion to 46.1\% in Hobbies \& Leisure 
(Table~\ref{tab:prevalence-category}).
\begin{table}[t]
\centering
\small
\setlength{\tabcolsep}{3pt}
\begin{tabular}{lrrrr}
\toprule
\textbf{Category} & \textbf{Queries} & \textbf{w/ AIO} & \textbf{Rate} & \textbf{\% of Corpus} \\
\midrule
Sports             & 28{,}446 & 2{,}782 &  9.8\% & 51.4\% \\
Entertainment      &  8{,}251 & 1{,}348 & 16.3\% & 14.9\% \\
Business \& Finance&  3{,}360 &   880  & 26.2\% &  6.1\% \\
Other              &  2{,}957 &   405  & 13.7\% &  5.3\% \\
Law \& Gov.        &  2{,}924 &   280  &  9.6\% &  5.3\% \\
Politics           &  2{,}166 &   162  &  7.5\% &  3.9\% \\
Climate            &  1{,}735 &   128  &  7.4\% &  3.1\% \\
Science            &  1{,}560 &   622  & 39.9\% &  2.8\% \\
Hobbies \& Leisure &  1{,}185 &   546  & 46.1\% &  2.1\% \\
Games              &    778  &    80  & 10.3\% &  1.4\% \\
Technology         &    464  &    86  & 18.5\% &  0.8\% \\
Food \& Drink      &    289  &    71  & 24.6\% &  0.5\% \\
Health             &    278  &    74  & 26.6\% &  0.5\% \\
Travel \& Transport&    265  &    23  &  8.7\% &  0.5\% \\
Jobs \& Education  &    252  &    43  & 17.1\% &  0.5\% \\
Shopping           &    158  &    23  & 14.6\% &  0.3\% \\
Beauty \& Fashion  &    142  &    5   &  3.5\% &  0.3\% \\
Autos \& Vehicles  &    138  &    19  & 13.8\% &  0.2\% \\
Pets \& Animals    &     45  &     6  & 13.3\% &  0.1\% \\
\midrule
\textbf{Total}     & \textbf{55{,}393} & \textbf{7{,}583} & \textbf{13.7\%} & \textbf{100.0\%} \\
\bottomrule
\end{tabular}
\caption{AIO activation rate by topical category, sorted by query volume.}
\label{tab:prevalence-category}
\end{table}
Politics (7.5\%) and Law \& Government (9.6\%) fall below average, consistent with Google's stated policy of exercising caution on sensitive topics~\cite{google_whitepaper_2025}, though Health (26.6\%) shows that this suppression is not applied uniformly across all sensitive areas.
The variation across categories reflects hidden editorial choices in AIO deployment that require continued tracking as the system continues to expand.
Importantly, activation rate and fidelity (Section~\ref{sec:eval-representativeness}) are largely independent  ($r \approx 0.313$; $p=0.192$), meaning that low-activation categories are not necessarily safer --- the proportion of unsupported claims within a category does not simply track how often AIOs appear there.

\paragraph{Query phrasing and length.}
Question-form queries\footnote{A query is classified as question-form if its leading whole word matches one of 15 interrogatives (\textit{who}, \textit{what}, \textit{where}, \textit{when}, \textit{why}, \textit{how}, \textit{which}, \textit{is}, \textit{are}, \textit{was}, \textit{can}, \textit{do}, \textit{does}, \textit{did}, \textit{has}).} trigger AIOs at 64.7\% versus 9.5\% for 
non-question queries, a 6.8 times difference 
($\chi^2 = 10{,}002.2$, $p < 0.001$).
Open-ended explanatory interrogatives activate most reliably, with \textit{how} at 84.3\% and \textit{why} at 73.4\%, while closed-form lookups activate less often, with \textit{who} at 47.9\% and \textit{did} at 39.8\%, though even the lowest interrogatives still trigger AIOs at least 4 times more than non-question queries (Table~\ref{tab:interrogative-activation}).
Query length independently amplifies activation: even restricting to non-question queries, the AIO rate climbs from 9.9\% for single-word queries to 38.7\% for queries of six or more words (Table~\ref{tab:prevalence-length}).
Taken together, both signals point in the same direction --- AIO exposure is concentrated among users posing open-ended questions that require synthesis and explanation, precisely the queries where the fidelity failures documented in Section~\ref{sec:eval-representativeness} are most consequential.

\begin{table}[t]
  \centering
  \small
  \begin{tabular}{l r r r}
    \toprule
    \textbf{Interrogative} & \textbf{AIOs} & \textbf{Queries} & \textbf{Rate} \\
    \midrule
    \emph{does}   &   12 &    12 & 100.0\% \\
    \emph{which}  &    2 &     2 & 100.0\% \\
    \emph{do}     &    1 &     1 & 100.0\% \\
    \emph{can}    &   34 &    35 &  97.1\% \\
    \emph{how}    &  451 &   535 &  84.3\% \\
    \emph{has}    &   70 &    87 &  80.5\% \\
    \emph{why}    &  102 &   139 &  73.4\% \\
    \emph{when}   &  313 &   431 &  72.6\% \\
    \emph{are}    &   49 &    72 &  68.1\% \\
    \emph{where}  &  715 & 1{,}092 &  65.5\% \\
    \emph{what}   &  402 &   663 &  60.6\% \\
    \emph{is}     &  296 &   533 &  55.5\% \\
    \emph{was}    &   16 &    32 &  50.0\% \\
    \emph{who}    &  169 &   353 &  47.9\% \\
    \emph{did}    &   84 &   211 &  39.8\% \\
    \midrule
    \textbf{All questions}    & \textbf{2{,}716} & \textbf{4{,}198} & \textbf{64.7\%} \\
    \textbf{All non-questions} & \textbf{4{,}867} & \textbf{51{,}195} & \textbf{9.5\%} \\
    \bottomrule
  \end{tabular}
    \caption{AIO activation rate by leading interrogative.}
  \label{tab:interrogative-activation}
\end{table}
\begin{table}[t]
\centering
\small
\setlength{\tabcolsep}{4pt}
\begin{tabular}{lrrrrrr}
\toprule
& \multicolumn{3}{c}{\textbf{All Queries}} & \multicolumn{3}{c}{\textbf{Non-Question Only}} \\
\cmidrule(lr){2-4} \cmidrule(lr){5-7}
\textbf{Words} & \textbf{$N$} & \textbf{AIOs} & \textbf{Rate} & \textbf{$N$} & \textbf{AIOs} & \textbf{Rate}\\
\midrule
1     &  5{,}949 &    593  & 10.0\% &  5{,}941 &    589  &  9.9\% \\
2     & 22{,}941 &    782  &  3.4\% & 22{,}939 &    781  &  3.4\% \\
3     & 12{,}918 &  1{,}678 & 13.0\% & 12{,}746 &  1{,}600 & 12.6\% \\
4     &  6{,}829 &  1{,}368 & 20.0\% &  6{,}305 &  1{,}066 & 17.0\% \\
5     &  3{,}148 &  1{,}066 & 33.9\% &  2{,}139 &    396  & 18.5\% \\
6+    &  3{,}608 &  2{,}096 & 58.1\% &  1{,}125 &    435  & 38.7\% \\
\bottomrule
\end{tabular}
\caption{AIO activation rate by query length (in words), overall and for non-question queries only.}
\label{tab:prevalence-length}
\end{table}

\subsection{Source Selection}
\label{sec:eval-sources}

Every AIO in our dataset embeds at least one explicit reference citation, making it possible to study not just whether an AIO appears but which sources Google elects to cite, how they compare in quality to co-displayed first-page results, and how much the two source pools overlap.
Throughout this section, we contrast AIO references against the first-page results shown on the same SERP --- two source pools selected by Google for the same query under different objectives.
We score domains using the PC1 metric introduced in Section~\ref{sec:doamin-quality}.
Note that during our observation window, publishers could restrict AIO use through general search controls such as \texttt{noindex} and \texttt{nosnippet} HTML/HTTP directives, which also affected traditional Search~\cite{google2026publishercontrols}.
Our source-distribution findings are therefore conditional on eligible content; publishers using these controls may be absent or underrepresented in the observed AIO citation pool.

\begin{figure}[t]
\centering
\includegraphics[width=0.9\columnwidth]{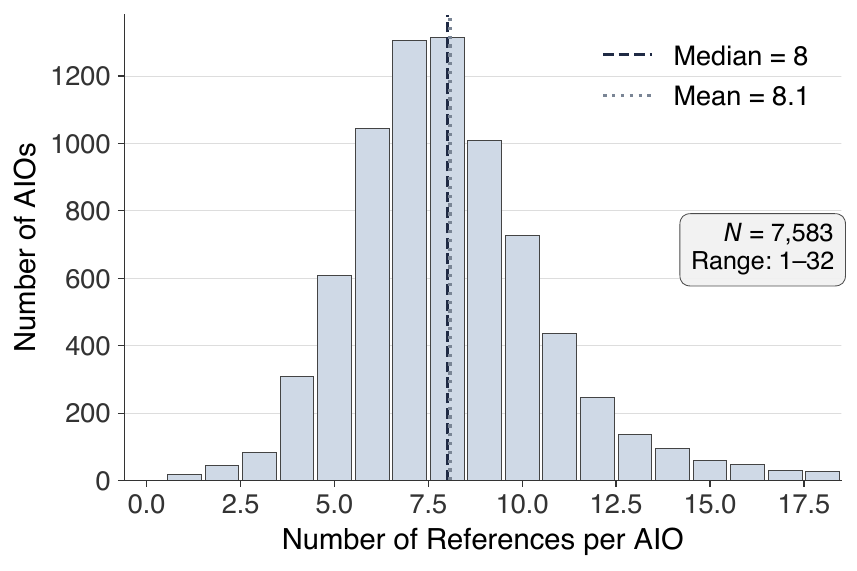}
\caption{Distribution of reference counts across 7{,}583 AIOs.}
\label{fig:ref-count-distribution_hist}
\end{figure}

\begin{figure}[t]
\centering
\includegraphics[width=0.9\columnwidth]{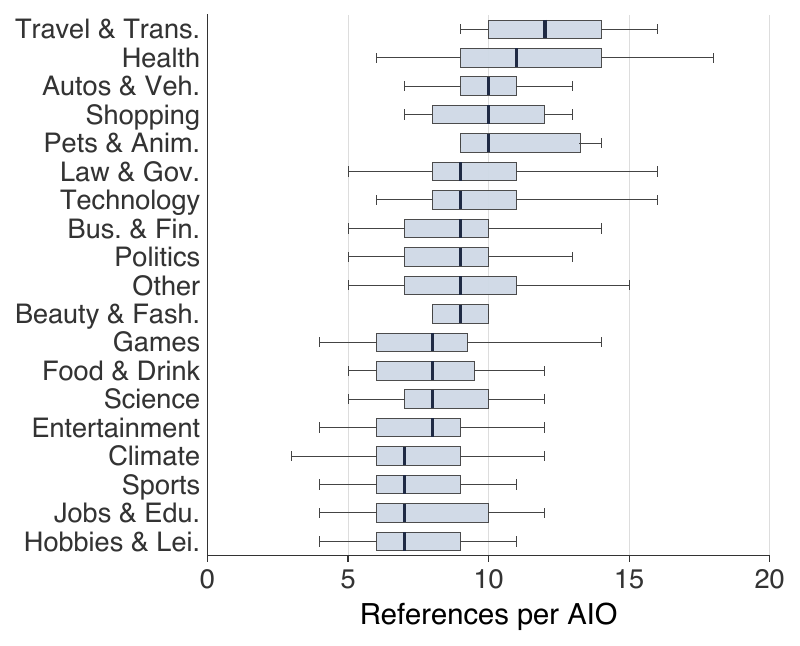}
\caption{Median reference count by topical category.}
\label{fig:ref-count-distribution_box}
\end{figure}

\paragraph{Reference reliance.}
Across the 7,583 AIOs, Google cited 61,212 reference URLs drawn from 7,479 unique hostnames, with a median of 8 references per AIO (mean 8.1, range 1--32; Figure~\ref{fig:ref-count-distribution_hist}).
Reference counts vary modestly across topical categories (Figure~\ref{fig:ref-count-distribution_box}): Travel \& Transportation (median 12), Health (median 11), and Shopping and Autos \& Vehicles (median 10) cite the most per AIO, while Sports, Jobs \& Education, Hobbies \& Leisure, and Climate cluster at the bottom (median 7).

The citation distribution is heavily long-tailed: the 5, 10, 50, and 100 most frequently cited hostnames account for 20.0\%, 29.7\%, 48.3\%, and 57.1\% of all AIO citations, respectively.
By comparison, first-page SERP results are markedly more concentrated: the 5, 10, 50, and 100 most frequently occurring hostnames account for 47.0\%, 54.5\%, 67.6\%, 72.3\% of its 241,710 SERP links, respectively.
The remaining 42.9\% of citations, outside the 100 most frequently cited hostnames, are distributed across 7,366 hostnames.
Of all 7,466 unique AIO hostnames in the comparison sample, 4,199 (56.2\%) are cited exactly once over the entire 40-day window, compared with 8,078 of 15,383 (52.5\%) unique SERP hostnames.
Source breadth, not concentration, is the dominant shape of AIO citation, in contrast to the more concentrated first-page SERP distribution. 
The five most frequently cited hostnames are \texttt{youtube.com} (5.49\%), \texttt{en.wikipedia.org} (4.39\%), \texttt{facebook.com} (3.68\%), \texttt{instagram.com} (3.65\%), and \texttt{usatoday.com} (2.80\%).
The presence of social and video platforms near the top of the distribution is notable; we examine their role as unvetted user-generated content in the UGC analysis below.

\paragraph{AIO vs.\ first-page source quality.}
AIO-cited domains are systematically more credible than co-displayed first-page results.
Across the 36,895 AIO references that resolve to a PC1 score, mean credibility is 0.732, compared with 0.611 across the 117,390 matched first-page URLs --- a gap of +0.122 on the $[0,1]$ scale (95\% CI $[0.120, 0.124]$; Welch's $t = 107.1$, $p < 0.001$).
This gap is not driven by a small number of high-volume categories.
%
\begin{table}[t]
  \centering
  \small
  \begin{tabular}{l rr rr l}
    \toprule
                          & \multicolumn{2}{c}{\textbf{AIO refs}} & \multicolumn{2}{c}{\textbf{First-page refs}} & \\
    \cmidrule(lr){2-3} \cmidrule(lr){4-5}
    Category              & PC1 & $n$ & PC1 & $n$ & $\Delta$ \\
    \midrule
    Science               & 0.769 &  3{,}896 & 0.633 & 12{,}175 & $+0.136$\textsuperscript{***} \\
    Business \& Finance   & 0.751 &  5{,}024 & 0.624 & 15{,}317 & $+0.127$\textsuperscript{***} \\
    Sports                & 0.748 & 11{,}943 & 0.611 & 37{,}497 & $+0.137$\textsuperscript{***} \\
    Hobbies \& Leisure    & 0.738 &  2{,}911 & 0.631 &  9{,}741 & $+0.107$\textsuperscript{***} \\
    Health                & 0.737 &      509 & 0.645 &  1{,}426 & $+0.091$\textsuperscript{***} \\
    Politics              & 0.737 &      874 & 0.636 &  2{,}838 & $+0.100$\textsuperscript{***} \\
    Food \& Drink         & 0.734 &      353 & 0.602 &      966 & $+0.132$\textsuperscript{***} \\
    Climate               & 0.717 &      559 & 0.622 &  1{,}919 & $+0.095$\textsuperscript{***} \\
    Law \& Government     & 0.714 &  1{,}354 & 0.617 &  4{,}142 & $+0.097$\textsuperscript{***} \\
    Other                 & 0.699 &  1{,}961 & 0.596 &  6{,}270 & $+0.104$\textsuperscript{***} \\
    Entertainment         & 0.695 &  6{,}445 & 0.588 & 21{,}078 & $+0.107$\textsuperscript{***} \\
    Jobs \& Education     & 0.687 &      130 & 0.582 &      556 & $+0.105$\textsuperscript{***} \\
    Pets \& Animals       & 0.656 &       27 & 0.526 &      111 & $+0.129$ \\
    Games                 & 0.652 &      218 & 0.523 &  1{,}029 & $+0.129$\textsuperscript{***} \\
    Technology            & 0.629 &      378 & 0.545 &  1{,}190 & $+0.083$\textsuperscript{***} \\
    Shopping              & 0.619 &      112 & 0.536 &      371 & $+0.083$\textsuperscript{***} \\
    Travel \& Transport.  & 0.603 &       98 & 0.512 &      345 & $+0.091$\textsuperscript{***} \\
    Autos \& Vehicles     & 0.586 &       82 & 0.526 &      349 & $+0.060$ \\
    Beauty \& Fashion     & 0.536 &       21 & 0.461 &       70 & $+0.076$ \\
    \midrule
    \textbf{All}          & \textbf{0.732} & \textbf{36{,}895} & \textbf{0.611} & \textbf{117{,}390} & $\mathbf{+0.122}$\textsuperscript{***} \\
    \bottomrule
  \end{tabular}
    \caption{Mean PC1 domain-quality score for AIO refs vs. organic results, by category, sorted by AIO descending. $n$ = URLs matched to PC1 (unmatched excluded; see Section~\ref{sec:doamin-quality}). Markers on $\Delta$: Welch's $t$-test, Bonferroni-corrected (\textsuperscript{***}~$p<0.001$, \textsuperscript{**}~$p<0.01$).}
  \label{tab:pc1-by-category}
\end{table}
Mean AIO PC1 ranges from 0.769 in Science and 0.751 in Business \& Finance at the top to 0.586 in Autos \& Vehicles and 0.536 in Beauty \& Fashion at the bottom (Table~\ref{tab:pc1-by-category}; one-way ANOVA: $F(18, 36{,}876)=58.0$, $p < 0.001$).
Per-category Welch's $t$-tests with Bonferroni correction reach significance in 16 of 19 categories, and all 19 category-level point estimates favor AIO references; no category shows a reversal.
The three non-significant categories --- Pets \& Animals, Autos \& Vehicles, and Beauty \& Fashion --- have small AIO subsamples ($n \leq 82$ matched references).
PC1 coverage differs between the two pools (60.4\% of AIO references vs.\ 48.6\% of first-page URLs), with unmatched entries dominated by social platforms and brand-owned domains that the PC1 release does not score (e.g., \url{nfl.com}).
Accordingly, the reported comparisons are conditional on coverage in the PC1 dataset; we do not impute credibility scores to unmatched domains.

As credibility measures differ in their construction and granularity, our results may depend on the choice of measure. 
Thus, we repeat the analysis using three alternative measures, i.e., MBFC factuality~\cite{mbfc,lin2023high}, Lasser mean~\cite{lasser,lin2023high}, and Burdisso~\cite{burdisso}. 
We find that the credibility gap favoring AIO references persists under MBFC and Lasser, whereas Burdisso's graph-based measure yields a small, non-significant difference in the opposite direction. 
We provide additional details in Appendix~\ref{appendix:credibility-sensitivity}.

\begin{table*}[t]
\centering
\small
\begin{tabular}{l r r r r r c}
\toprule
SERP specification
& Paired $n$
& AIO PC1
& First-page PC1
& $\Delta$
& $d_z$
& 95\% CI \\
\midrule
Top 1
& 5{,}020
& 0.740
& 0.777
& $-0.037$
& $-0.235$
& $[-0.042,-0.033]$ \\
Top 3
& 6{,}534
& 0.736
& 0.753
& $-0.017$& $-0.117$& $[-0.020,-0.013]$ \\
Top 5
& 6{,}942
& 0.734
& 0.740
& $-0.006$
& $-0.045$
& $[-0.009,-0.003]$ \\
Top 10
& 7{,}342
& 0.730
& 0.696
& $+0.034$
& $+0.264$
& $[0.031,0.037]$ \\
Full page
& 7{,}444
& 0.728
& 0.618
& $+0.110$
& $+0.978$
& $[0.108,0.113]$ \\
\midrule
Geometric ($\rho=0.5$)
& 7{,}444
& 0.728
& 0.744
& $-0.016$
& $-0.111$
& $[-0.019,-0.013]$ \\
Geometric ($\rho=0.8$)
& 7{,}444
& 0.728
& 0.707
& $+0.021$
& $+0.179$
& $[0.019,0.024]$ \\
\bottomrule
  \end{tabular}
  \caption{Query-paired mean PC1 under top-$k$ and geometric SERP specifications. $\Delta$ and Cohen's $d_z$ denote AIO minus first-page PC1; top-$k$ follows the stored cleaned SERP order.}
  \label{tab:pc1-rank-sensitive}
\end{table*}

\paragraph{Source quality by search result ranking.}
The preceding comparison weights all links with PC1 scores equally, although users may pay more attention to results near the top of the SERP.
Thus, we repeat the analysis within each query, comparing mean PC1 across AIO references with that of the first $k\in{1,3,5,10}$ links and the full first page.
Each comparison includes only URLs with PC1 scores and queries with at least one such URL on both sides; the eligible queries therefore vary across comparisons.
We also weight first-page links by position using $w_r=(1-\rho)\rho^{r-1}$, normalized over links with PC1 scores within each query.
With $\rho=0.5$, the weight halves at each successive position, whereas with $\rho=0.8$, it decreases by 20
For example, the first three positions receive relative weights of $1,0.5,0.25$ under $\rho=0.5$, compared with $1,0.8,0.64$ under $\rho=0.8$.

We find that first-page results have higher mean credibility than AIO references when considering the first link or the first three links (Table~\ref{tab:pc1-rank-sensitive}).
The difference is small when considering the first five links ($\Delta=-0.006$).
In contrast, AIO references have higher mean credibility when considering the first ten links ($\Delta=+0.034$) or all first-page links ($\Delta=+0.110$).
We also find that first-page results have higher mean credibility when weights decrease more rapidly with rank ($\rho=0.5$; $\Delta=-0.016$).
However, AIO references have higher mean credibility when weights decrease more slowly ($\rho=0.8$; $\Delta=+0.021$).
All seven comparisons are statistically significant after Bonferroni correction.
However, we note that the effect sizes are small or negligible for all comparisons except when considering all first-page links, where the effect size is large (Cohen's $d_z=0.978$).

Overall, our unweighted full-page comparison shows that AIO references have higher mean credibility than first-page results. 
However, we find that this advantage becomes small or reverses when we consider earlier first-page links or give them greater weight. 
These results indicate that AIO references are not consistently more credible than first-page results; the credibility gap depends on which links are considered and how they are weighted.

\paragraph{UGC prevalence.}
We define UGC as content from platforms without strong editorial moderation: \texttt{facebook.com}, \texttt{instagram.com}, \texttt{linkedin.com}, \texttt{pinterest.com}, \texttt{quora.com}, \texttt{reddit.com}, \texttt{threads.com}, \texttt{tiktok.com}, \texttt{twitter.com}, \texttt{x.com}, and \texttt{youtube.com}. We exclude \texttt{wikipedia.org} because its content is governed through community-based editorial policies and norms~\cite{wikipedia_governance}.
14.2\% of AIO reference URLs (8,671 of 61,046) come from these platforms, compared with 49.9\% of first-page result URLs (120,695 of 241,710).
%
\begin{table}[t]
  \centering
  \small
  \resizebox{\columnwidth}{!}{%
  \begin{tabular}{l rr rr l}
    \toprule
                          & \multicolumn{2}{c}{\textbf{AIO refs}} & \multicolumn{2}{c}{\textbf{First-page refs}} & \\
    \cmidrule(lr){2-3} \cmidrule(lr){4-5}
    Category              & UGC \% & $n$ & UGC \% & $n$ & $\Delta$ \\
    \midrule
    Beauty \& Fashion     & 28.85 &      52  & 60.99 &       223 & $-32.14$\textsuperscript{**}  \\
    Autos \& Vehicles     & 27.42 &      186 & 46.87 &       958 & $-19.45$ \\
    Shopping              & 25.86 &      232 & 54.41 &       952 & $-28.55$\textsuperscript{***} \\
    Pets \& Animals       & 20.90 &       67 & 47.19 &       320 & $-26.29$\textsuperscript{***}  \\
    Technology            & 20.29 &      828 & 53.37 &   3{,}013 & $-33.08$\textsuperscript{***} \\
    Travel \& Transport.  & 20.00 &      290 & 45.59 &   1{,}303 & $-25.59$\textsuperscript{***} \\
    Entertainment         & 19.41 & 10{,}461 & 56.63 &  44{,}022 & $-37.22$\textsuperscript{***} \\
    Food \& Drink         & 18.72 &      561 & 52.34 &   2{,}117 & $-33.62$\textsuperscript{***} \\
    Games                 & 18.13 &      662 & 53.82 &   2{,}863 & $-35.70$\textsuperscript{***} \\
    Other                 & 16.21 &  3{,}725 & 44.93 &  14{,}018 & $-28.71$\textsuperscript{***} \\
    Science               & 15.33 &  5{,}204 & 56.22 &  21{,}068 & $-40.88$\textsuperscript{***} \\
    Law \& Government     & 14.55 &  2{,}708 & 37.77 &   9{,}406 & $-23.22$\textsuperscript{***} \\
    Politics              & 13.93 &  1{,}436 & 39.02 &   5{,}484 & $-25.10$\textsuperscript{***} \\
    Business \& Finance   & 13.37 &  7{,}919 & 49.42 &  30{,}003 & $-36.04$\textsuperscript{***} \\
    Jobs \& Education     & 12.24 &      343 & 40.79 &   1{,}466 & $-28.55$\textsuperscript{***} \\
    Hobbies \& Leisure    & 11.68 &  4{,}040 & 44.92 &  17{,}839 & $-33.24$\textsuperscript{***} \\
    Sports                & 11.20 & 20{,}557 & 50.32 &  80{,}023 & $-39.11$\textsuperscript{***} \\
    Health                & 10.73 &      839 & 26.68 &   2{,}729 & $-15.95$\textsuperscript{***} \\
    Climate               &  9.29 &      936 & 35.74 &   3{,}903 & $-26.45$\textsuperscript{***} \\
    \midrule
    \textbf{All}          & \textbf{14.20} & \textbf{61{,}046} & \textbf{49.93} & \textbf{241{,}710} & $\mathbf{-35.73}$\textsuperscript{***} \\
    \bottomrule
  \end{tabular}%
  }
    \caption{Share of unvetted UGC URLs in AIO refs vs. organic results, by category, sorted by AIO descending. $n$ = total URLs in category. Markers on $\Delta$: $\chi^{2}$ test, Bonferroni-corrected (\textsuperscript{***}~$p<0.001$, \textsuperscript{**}~$p<0.01$).}
  \label{tab:ugc-by-category}
\end{table}
Within AIO citations, four platforms account for 96.4\% of the UGC contribution: \texttt{youtube.com} (5.49\% of AIO citations), \texttt{facebook.com} (3.67\%), \texttt{instagram.com} (3.66\%), and \texttt{reddit.com} (0.87\%); all other UGC platforms together contribute for 0.51\%.
 
The AIO UGC share varies by a factor of 3.1 across categories, from 9.3\% in Climate and 10.7\% in Health to 28.9\% in Beauty \& Fashion and 27.4\% in Autos \& Vehicles (Table~\ref{tab:ugc-by-category}).
This variation reflects the degree to which those topics are naturally discussed through first-hand consumer experience and visual content on social platforms rather than through institutional sources.
Across all 19 categories, AIOs cite UGC significantly less often than the corresponding cleaned first-page link pools after Bonferroni correction. The gaps range from $-15.9$ percentage points in Health to $-40.9$ percentage points in Science, where 56.22\% of first-page links come from the defined UGC platforms, compared with 15.33\% of AIO references.
This consistent pattern partly explains the higher average credibility of AIO-cited sources.

\paragraph{Reference overlap with first-page results.}
AIO source selection is not simply a re-ranking of the first page.
Averaged across the 7,562 AIOs, only 31.5\% of AIO reference domains overlap with the top-5 first-page results, rising to 48.0\% at top-10 and 70.2\% across the full first page.
In fact, 29.8\% of AIO reference domains do not appear anywhere on the corresponding first page.
These off-page references are not lower quality.
Among PC1-covered references, mean PC1 is 0.757 for the 9,218 AIO references whose hosts are absent from the SERP (off-page references), compared with 0.724 for the 27,677 on-page references ($\Delta\text{PC1}=+0.033$, $p<0.001$).
Using all AIO references, the corresponding UGC shares are 3.39\% (590 of 17,399) among off-page references and 18.52\% (8,081 of 43,641) among on-page references ($\Delta\text{UGC}=-15.13$ percentage points, $p<0.001$).
Google describes AIOs as grounded in its core ranking infrastructure~\cite{google_whitepaper_2025}, but the AIO and first-page mechanisms clearly apply different selection criteria even when acting on the same index.
Thus, publishers whose content informs an AIO are not necessarily the same publishers whose pages users would encounter by scrolling past it.

\subsection{Claim Fidelity}
\label{sec:eval-representativeness}

We evaluate how faithfully AIOs represent the information in their cited sources by decomposing each AIO into atomic factual claims and verifying each against the full text of every cited reference.
Following the pipeline described in Section~\ref{sec:methodology-representativeness}, we produce 98,098 claims across 7,583 AIOs (mean 12.9 per AIO, median 12, range 0--64).
Of these AIOs, 7,491 (98.8\%) yield at least one extracted claim and at least one cited reference with scrapable body text. The remaining 92 are excluded (80 yielded zero claims, 12 cite only social-media URLs whose bodies we do not extract).
Across the 7,491 verifiable AIOs, the pipeline returns 98,020 claim-level judgments, each assigned one of five labels --- \textsc{Clear}, \textsc{Vague}, \textsc{Ambiguous}, \textsc{Incorrect}, or \textsc{Omitted} (Table~\ref{label_examples}).  For some analyses, we collapse into \emph{consistent} (\textsc{Clear} + \textsc{Vague}) versus \emph{inconsistent} (\textsc{Ambiguous} + \textsc{Incorrect} + \textsc{Omitted}) claims.

\paragraph{Overall fidelity.}

Of the 98,020 verified claims, 87,204 (88.97\%) are consistent: 84.61\% are \textsc{Clear} and 4.36\% are \textsc{Vague}.
The remaining 11.03\% are split across three failure modes: \textsc{Omitted} at 6.98\% (no cited source mentions the claim at all), \textsc{Incorrect} at 2.66\% (a cited source explicitly contradicts the claim), and \textsc{Ambiguous} at 1.39\% (cited sources disagree with each other; Table~\ref{tab:verification_label_distribution}).
At the AIO level, the median AIO has 93.33\% of its claims grounded; 3,141 of the 7,491 verified AIOs (41.9\%) are perfectly grounded, and 4,631 (61.8\%) are at least 90\% consistent.
The failure tail, while small, is non-trivial: 205 AIOs (2.74\%) have fewer than half their claims grounded, and 64 (0.85\%) have zero consistent claims.

Two patterns in the failure modes are worth highlighting.
First, omission dominates active misrepresentation --- when an AIO claim is not grounded in its cited references, it is roughly $2.6$ times more likely to be a fact no cited source mentions than a fact a cited source contradicts.
A non-trivial share of these inconsistencies, however, may be a measurement artifact of our pipeline rather than a genuine AIO failure as we exclude social and video platforms (Section~\ref{sec:webpage_content}).
To bound this artifact, we examined \texttt{Inconsistent} claims and identified those appearing in AIOs that cite at least one UGC source.
We found that 59.9\% of \texttt{Inconsistent} claims (5,658 of 9,449) meet this criterion (54.7\% of \texttt{Incorrect}, 61.8\% of \texttt{Omitted}).
Even under the most generous assumption (i.e., that every such claim is in fact supported by its uncrawled UGC source) the residual inconsistency rate would fall from 11.0\% to roughly 5.3\%, leaving a substantive floor of unsupported claims that cannot be attributed to UGC exclusion.
Second, the \textsc{Incorrect} category, though smaller in volume (2,609 claims), represents direct factual conflicts between an AIO's assertion and a source the AIO itself cites in support of that assertion --- a failure mode a reader who sees only the AIO summary has no way to detect.

\begin{figure}[t]
\centering
\includegraphics[width=\columnwidth]{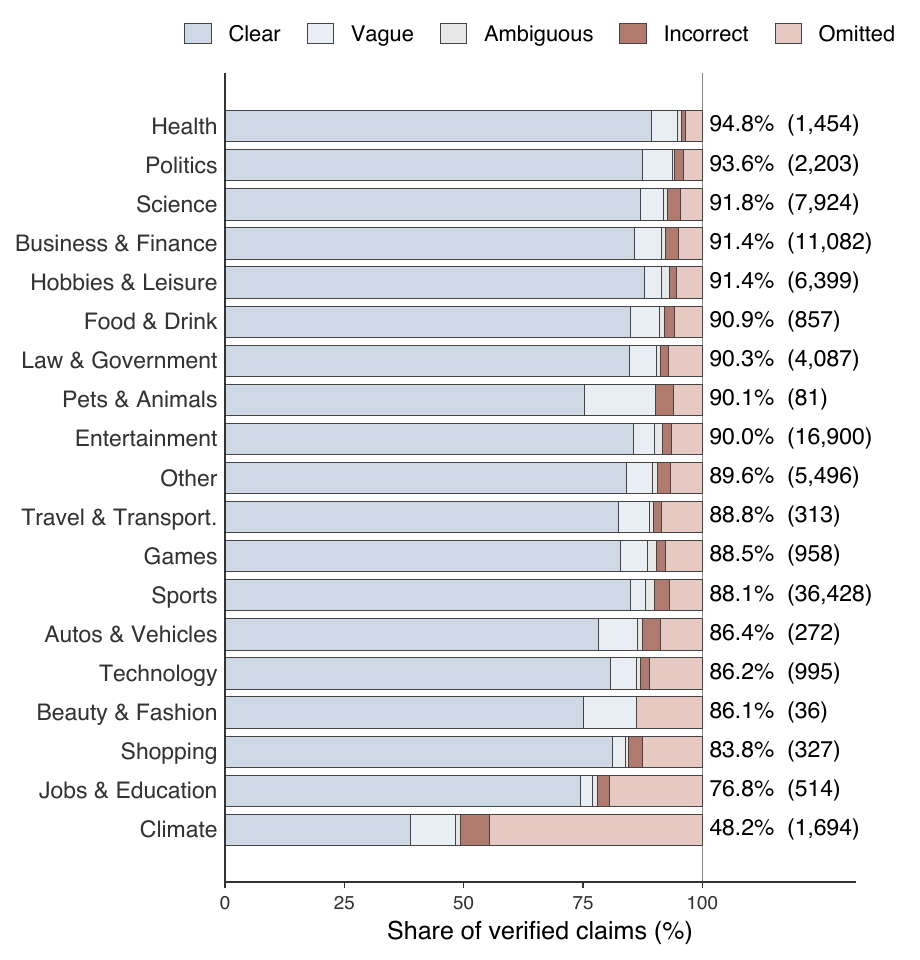}
\caption{Verification label distribution by topic category,
sorted by consistent share (\textsc{Clear}~$+$~\textsc{Vague})
descending; number of claims in parentheses.}
\label{fig:fidelity-by-category}
\end{figure}

\paragraph{Variation across topics.}
The 88.97\% aggregate consistent rate masks substantial variation across categories (Figure~\ref{fig:fidelity-by-category}).
Excluding Climate, whose anomalously low rate is a measurement artifact discussed below, per-category fidelity ranges from 76.85\% in Jobs \& Education to 94.77\% in Health.
The highest-fidelity categories are Health (94.77\%), Politics (93.65\%), Science (91.82\%), Business \& Finance (91.42\%), and Hobbies \& Leisure (91.40\%). 
It is worth noting that several of the highest-fidelity categories are also among the most consequential for users --- topics where inaccurate information carries real risk, and where Google may apply stricter grounding requirements consistent with its own YMYL (Your Money or Your Life) content policy~\cite{google_whitepaper_2025}.

The lower-fidelity categories warrant closer inspection.
Three of the four lowest non-Climate categories contain small but identifiable subsets of real-time queries --- school closings in Jobs \& Education, weather in Technology, and trending product queries in Shopping.
Excluding these sub-populations narrows each category's gap substantially: Technology rises to 89.41\%, Shopping to 85.90\%, and Jobs \& Education to 87.71\%. 
After this normalization, 18 of 19 categories sit within an 85.9\%--94.8\% band.

\paragraph{The Climate and Jobs anomalies.}
Climate and Jobs \& Education, the two lowest-consistency categories in Figure~\ref{fig:fidelity-by-category}, share a pipeline-level issue rather than a genuine AIO quality problem.
Climate's 48.23\% rate is dominated by weather and forecast queries (``weather today,'' ``forecast,'' ``pollen count today'') for which the AIO surfaces specific values --- temperatures, precipitation chances, wind speeds --- drawn from Google's structured weather feeds rather than the textual body of the cited pages.
Those pages (\texttt{weather.com}, \texttt{forecast.weather.gov}) update continuously; by the time our crawler captures them, often hours or days after the AIO was synthesized, the figures have shifted, and the verifier correctly labels the AIO claim \textsc{Omitted} or \textsc{Incorrect} even though Google was accurately reporting a real-time value at the time of generation.
Jobs \& Education exhibits the same pattern at smaller scale on school-closure queries (``school closings for tomorrow,'' ``marion county,'' ``bridgeport''), where district websites publish updates minute-by-minute during weather events.
In both cases Google may be correctly synthesizing values from a structured source whose textual representation our pipeline cannot freeze in time.
The 9.64\% combined \textsc{Omitted} and \textsc{Incorrect} rate should therefore be read as a conservative ceiling on substantive AIO unfaithfulness rather than a precise estimate. 
Appendix~\ref{app:realtime_queries} describe our approach to identify real-time categories across all categories.

\subsection{Economic Impact}
\label{sec:eval-impact}

AIOs consume publisher content as input and substitute a synthesized answer for the click-through that would normally have reached the publisher's page.
We do not measure traffic loss directly, but we do measure the monetization characteristics of the pages AIOs draw on, and whether Google's own ad inventory is preserved on AIO-bearing pages.

\paragraph{Ad prevalence on cited pages.}
Of the 61,212 reference URLs AIOs cite over the 40-day window, 30,994 (50.63\%) display visible ads --- meaning more than half the pages whose content makes AIO synthesis possible are running ad-supported business models that depend on the page-view traffic AIOs intercept.
This figure is a conservative lower bound.
A further 14.2\% of references point to social and video platforms --- including \texttt{facebook.com}, \texttt{instagram.com}, \texttt{reddit.com}, and \texttt{tiktok.com} --- whose pages our pipeline does not crawl and which we treat as carrying no ads, even though in practice these platforms run large ad businesses.
Thus, the true ad-supported share of the AIO citation corpus is 
meaningfully higher than what we directly observe.

\paragraph{Variation by topical category.}
Publisher monetization is uneven across the topics AIOs draw on 
(Table~\ref{tab:economic-by-category}).
\begin{table}[t]
  \centering
  \small
  \begin{tabular}{l rr r}
    \toprule
    Category              & Refs    & UGC     & Ads \% \\
    \midrule
    Sports                & 20{,}595 & 2{,}306 & 60.19 \\
    Entertainment         & 10{,}516 & 2{,}039 & 55.50 \\
    Business \& Finance   &  7{,}944 & 1{,}062 & 39.97 \\
    Science               &  5{,}212 &    798  & 41.85 \\
    Hobbies \& Leisure    &  4{,}067 &    475  & 63.14 \\
    Other                 &  3{,}733 &    604  & 40.05 \\
    Law \& Government     &  2{,}708 &    394  & 29.47 \\
    Politics              &  1{,}436 &    200  & 36.56 \\
    Climate               &     941  &     87  & 48.46 \\
    Health                &     839  &     90  & 27.77 \\
    Technology            &     828  &    168  & 42.75 \\
    Games                 &     662  &    120  & 41.54 \\
    Food \& Drink         &     561  &    105  & 57.93 \\
    Jobs \& Education     &     343  &     42  & 34.99 \\
    Travel \& Transport.  &     290  &     58  & 27.24 \\
    Shopping              &     232  &     60  & 35.34 \\
    Autos \& Vehicles     &     186  &     51  & 28.49 \\
    Pets \& Animals       &      67  &     14  & 32.84 \\
    Beauty \& Fashion     &      52  &     15  & 38.46 \\
    \midrule
    \textbf{All}          & \textbf{61{,}212} & \textbf{8{,}688} & \textbf{50.63} \\
    \bottomrule
  \end{tabular}
  \caption{Ad prevalence on AIO-cited pages, by category,
    sorted by reference volume. UGC refs are not
    scrape-analyzed; reported rates are lower bounds}
  \label{tab:economic-by-category}
\end{table}
Volume-weighted, publisher-side exposure to AIO displacement is concentrated in a few categories: Sports alone accounts for 33.6\% of all AIO references, followed by Entertainment (17.2\%) and Business \& Finance (13.0\%).
Low observed ad rates in Health, Law \& Government, and Travel \& Transportation (27--30\% ad rates) do not signal that these categories are safe from displacement --- many high-quality sites in these areas monetize through subscriptions, paywalls, or direct conversions rather than display advertising, and those revenue paths are equally defeated when an AIO answers the query without a click.

\paragraph{Sponsored search ads on AIO-bearing pages.}
Of the 7,583 AIO-bearing SERPs in our window, 164 (2.16\%) also carry at least one Google sponsored search ad on the same page, and 39 (0.51\%) place a sponsored slot above the AIO block.
No sponsored slots appear inside the AIO container itself.
When AIO and sponsored ads coexist they occupy disjoint page regions, meaning AIO is additive to Google's ad inventory rather than substitutive. Google's sponsored-search revenue capture is preserved even as AIO substitutes for the click that would have reached a publisher's page.
The 2.16\% co-occurrence rate reflects our query sample, which is dominated by informational and trending queries rather than commercial-intent searches where sponsored ads are most prevalent.
The co-occurrence cases we do observe are concentrated in Sports (42 cases), Business \& Finance (34), Entertainment (16), Politics (13), and Law \& Government (11).
The asymmetry we document --- AIOs restructure the page in a way that preserves Google's ad capture while reducing click-throughs to the publishers whose content makes the AIO possible --- likely carries greater force in the commercial-intent query space our methodology underweights.
\section{Concluding Remarks}

We begin by noting that several of our findings reflect genuine quality investments on Google's part.
AIO-cited sources are systematically more credible than co-displayed first-page results, UGC is suppressed relative to organic ranking, and fidelity is highest in the categories --- Health, Politics, Science --- where errors carry the greatest consequence.
These patterns suggest deliberate design choices rather than incidental outcomes, and they are worth acknowledging before turning to the problems.

\paragraph{Generative AI Limitations}
The most fundamental of those problems is not one Google can fully solve.
Hallucination and unsupported claims are inherent to generative AI at its current state of development~\cite{huang2025hallucination,openai2025whyhallucinate}, and adding high-quality or better-curated inputs does not make them disappear~\cite{magesh2025hallucinations,wu_2025}.
Our findings, as well as those of others, show that unsupported claims can still be identified with targeted efforts.
What can be done is to continue driving down the current 11\% unsupported rate (Section \ref{sec:eval-representativeness}) and to more carefully limit deployment in contexts where errors are likely or carry significant consequences.
At the scale Google AIOs operate, even modest improvements in fidelity translate into large absolute reductions in the number of unsupported claims users encounter daily, but achieving that will require more than just better sources.

\paragraph{Economic Consequences}
Beyond fidelity, the economic consequences of AIO deployment deserve more attention than they have received.
More than half the pages AIOs cite depend on advertising revenue that evaporates when users do not click through, and the causal evidence now establishes that AIOs do reduce click-throughs substantially~\cite{agarwal2026field,pew2025_ai_clicks}.
The concern is not simply that individual publishers lose revenue today, but that sustained traffic displacement disincentivizes the production of the high-quality content that currently grounds the most faithful AIO responses.
This creates a feedback loop worth watching: degraded sourcing options lead to greater reliance on user-generated content, which leads to lower fidelity, which further erodes the incentive to invest in original reporting and research.
Potential remedies include revenue-sharing or licensing arrangements between Google and content producers, design choices that make citations more prominent and click-throughs easier, and deployment restrictions in topic categories where the content ecosystem is most financially fragile.
None of these is a complete solution, and all involve tradeoffs that Google is better positioned to evaluate than outside researchers.
What is clear is that the current arrangement --- in which publishers bear the costs of a system they did not choose and cannot meaningfully opt out of without consequences~\cite{google2026publishercontrols} --- is unlikely to remain unaddressed.
Our data offer one concrete measure of what is at stake. A large percentage of AIO-cited pages carry display advertising, and with Google reporting over 2 billion monthly AIO users~\cite{pichai2025alphabetq2earnings}, even small per-query reductions in click-through rates aggregate to economically significant traffic losses across the publisher ecosystem.
The regulatory proceedings already underway in the United Kingdom, the European Union, and the United States suggest that if Google does not act, policymakers might~\cite{cma2026,ec2025,doj2025}.

\paragraph{Transparency \& Accountability}
A third set of concerns involves transparency and accountability.
Better system design would include clearer public policies about when AIOs activate and why, measurable performance targets published by category, and genuine support for independent auditing.
Our study required substantial engineering effort to produce a single 40-day snapshot of a system that serves billions of queries daily.
Audit-ready APIs or structured data-sharing agreements with independent researchers would make this kind of measurement routine rather than exceptional, and would allow the field to track how the system evolves as Google continues to expand and modify it.

Finally, several questions this study cannot answer should be priorities for future work.
As our sampling is conditioned on queries receiving elevated or growing public attention, it may not reflect the full distribution of Google Search traffic, particularly less time-sensitive or long-tail queries.
We measure what AIOs say and what they cite, but not what users do with them --- whether errors propagate into belief~\cite{xu2025aisummaries}, whether corrections are sought, and whether the effects are more serious for high-stakes topics like health and political information.
We also do not know how AIO behavior varies across user profiles or personalization states. 
Our activation findings suggest that question-form queries are disproportionately exposed to AIO responses, and there is reason to think that users who phrase queries as natural language questions skew toward those less familiar with a topic and more reliant on the answer they receive.
Design choices that seem neutral --- triggering AIOs on question-form queries, for instance --- may not be neutral in their distributional effects across user populations, and understanding these heterogeneous effects is important both for evaluating the system's social consequences and for identifying the populations most at risk.

\paragraph{Takeaway}
Taken together, our findings describe a system that is performing better than its critics often claim and worse than Google's own public characterizations suggest.
AIOs cite credible sources, but nearly one in nine claims is unsupported.
They activate selectively, but the selection logic is opaque and concentrated among users asking open-ended questions.
Their source selection operates independently of Google's own ranking algorithm, raising questions about accountability for what appears in the answer.
And the publishers whose content makes the system possible are bearing economic costs that Google's own ad business does not share.
As AIOs become the default interface through which billions of people encounter information about the world, these are not merely technical questions --- they are questions about the design of a core piece of public information infrastructure, and they deserve sustained independent scrutiny.

\begin{acks}
The authors would like to thank the reviewers for their valuable feedback. This work was partially supported by the NSF (CNS-2543517).
\end{acks}
\bibliographystyle{ACM-Reference-Format}
\bibliography{refs}

\appendix
\section{Ethics}
\label{sec:ethics}

Our study involves automated web interactions that may have unintended side effects.
We outline these considerations and the steps we took to minimize potential impact.

First, our queries to Google trigger AIOs, which can incur computational costs for the provider. 
However, because we focus on trending queries, some responses may already be cached by the LLM hosting infrastructure, a practice that is becoming increasingly common~\cite{kwon2023efficient}.
Second, our methodology includes visiting webpages to determine whether they display advertisements. 
These visits may generate ad impressions and could, in principle, affect advertiser metrics or revenue. 
To mitigate this impact, we deliberately avoided clicking on ads or interacting with them in any way. 
This concern is not unique to our work, and in fact, it is common across web measurement studies.
Finally, in collecting web content, we adhered to access restrictions. 
We did not attempt to bypass paywalls, nor did we employ workarounds to scrape platforms that explicitly prohibit automated access (e.g., social media sites such as Reddit).
\section{Sensitivity to Alternative Credibility Measures}
\label{appendix:credibility-sensitivity}

Our findings may depend on the choice of credibility measure.
We therefore repeat the analysis using three alternatives: MBFC factuality~\cite{mbfc,lin2023high}, Lasser mean~\cite{lasser,lin2023high}, and Burdisso \texttt{p-reliability}~\cite{burdisso}.
These measures differ in their construction and granularity.
MBFC factuality and Lasser mean are coarser ordinal measures.
Specifically, they rank domains from lower to higher credibility using a fixed set of score levels. As a result, they cannot distinguish among domains assigned the same level.
Burdisso \texttt{p-reliability}, in contrast, estimates reliability from hyperlinks among news domains and produces a continuous score on $[-1,1]$.
Unlike the first two measures, it is not restricted to a fixed set of score levels.
We analyze each measure on the numeric scale provided in its original release.

To isolate the effect of measure choice from differences in domain coverage, we compare PC1 with each alternative measure on a shared sample.
For each comparison, we retain only the domain/path keys covered by both measures and evaluate both measures on identical URL occurrences.
Within each search query, we separately average the scores of AIO references and first-page SERP links covered by both measures, compare the resulting means, and conduct a paired test at the query level.

Table~\ref{tab:pc1-sensitivity} presents the results across different measures. 
We note that the credibility gap favoring AIO references is reproduced by MBFC and Lasser, whereas Burdisso yields a small, non-significant gap favoring first-page results.
On the PC1--MBFC shared sample, mean query-level credibility is higher for AIO references under both PC1 (gap of $+0.133$) and MBFC (gap of $+0.106$, $p<0.001$).
On the PC1--Lasser shared sample, mean query-level credibility is also higher for AIO references under both PC1 (gap of $+0.008$) and Lasser (gap of $+0.013$, $p<0.001$).
On the PC1--Burdisso shared sample, PC1 shows slightly higher mean credibility for AIO references (gap of $+0.005$), whereas Burdisso shows slightly lower mean credibility for AIO references (gap of $-0.007$).
However, the confidence interval for the Burdisso gap includes zero ($p=0.130$), providing no evidence of a statistically significant difference in either direction.

Note that MBFC and Lasser are derived from rating sources used in the construction of PC1 and therefore do not provide fully independent checks.
Burdisso, in contrast, estimates reliability using hyperlink relationships among news domains and provides a methodologically distinct comparison.
Overall, the credibility gap favoring AIO references persists under the two rating-based alternatives but is not reproduced by the graph-based Burdisso measure.

\begin{table*}[t]
  \centering
  \small
  \setlength{\tabcolsep}{3pt}
  \renewcommand{\arraystretch}{1.08}
  \begin{tabular*}{\textwidth}{@{\extracolsep{\fill}}llrrrrr@{}}
    \toprule
    Comparison & Measure & $n$ & AIO mean & SERP mean & Gap [95\% CI] & $p$ \\
    \midrule
    \multirow{2}{*}{PC1 \& MBFC}
      & PC1 & \multirow{2}{*}{7,281}
      & 0.742 & 0.608 & $+0.133$ [$+0.130$, $+0.136$] & $<0.001$ \\
      & MBFC factuality &
      & 0.696 & 0.590 & $+0.106$ [$+0.103$, $+0.109$] & $<0.001$ \\
    \midrule
    \multirow{2}{*}{PC1 \& Lasser}
      & PC1 & \multirow{2}{*}{6,834}
      & 0.835 & 0.827 & $+0.008$ [$+0.006$, $+0.009$] & $<0.001$ \\
      & Lasser mean &
      & 0.886 & 0.874 & $+0.013$ [$+0.011$, $+0.015$] & $<0.001$ \\
    \midrule
    \multirow{2}{*}{PC1 \& Burdisso}
      & PC1 & \multirow{2}{*}{7,174}
      & 0.791 & 0.786 & $+0.005$ [$+0.004$, $+0.007$] & $<0.001$ \\
      & Burdisso \texttt{p-reliability} &
      & 0.669 & 0.677 & $-0.007$ [$-0.017$, $+0.002$] & $0.130$ \\
    \bottomrule
  \end{tabular*}
  \caption{Query-paired sensitivity to alternative credibility measures. The comparison column identifies the two measures evaluated on the same queries and URL occurrences. The $n$ column reports the number of queries for which both the AIO references and first-page results contain at least one URL covered by both measures. The gap is the AIO mean minus the SERP mean; 95\% confidence intervals and two-sided $p$-values are based on paired query-level differences.}
  \label{tab:pc1-sensitivity}
\end{table*}

\section{Claim Extraction Prompt}
\label{appendix:claim-extraction-prompt}

This appendix contains the full prompt used for atomic claim extraction (Section~\ref{sec:claim-extraction}).

\subsection{Claim Extraction Prompt}
Listing~\ref{lst:claim-extraction-prompt} presents the prompt used for claim extraction from AIOs.
Below we present an example AIO and claims extracted from that AIO. 

\subsubsection{AIO}
\textit{The 2026 NFL Draft will be held from April 23–25, 2026, in Pittsburgh, Pennsylvania, with main events at Point State Park and Acrisure Stadium. The three-day event begins with the first round on Thursday night, followed by rounds 2–7 over the next two days. 2026 NFL Draft Schedule (All Times ET): Round 1: Thursday, April 23, 8 p.m. Rounds 2-3: Friday, April 24, 7 p.m. Rounds 4-7: Saturday, April 25, Noon How to Watch:The draft will be broadcast on ESPN, ABC, NFL Network, and ESPN Deportes, with streaming options available on NFL+ and the ESPN App. ESPN ABC NFL Network ESPN Deportes NFL+ ESPN App}

\subsubsection{Extracted Claims}
\begin{enumerate}[leftmargin=15pt]
    \item The 2026 NFL Draft will be held from April 23–25, 2026, in Pittsburgh, Pennsylvania.
\item Main events of the 2026 NFL Draft will be at Point State Park and Acrisure Stadium.
\item The 2026 NFL Draft is a three-day event.
\item The first round of the 2026 NFL Draft begins on Thursday night.
\item Rounds 2–7 of the 2026 NFL Draft will occur over the next two days.
\item Round 1 of the 2026 NFL Draft is scheduled for Thursday, April 23, at 8 p.m. ET.
\item Rounds 2-3 of the 2026 NFL Draft are scheduled for Friday, April 24, at 7 p.m. ET.
\item Rounds 4-7 of the 2026 NFL Draft are scheduled for Saturday, April 25, at Noon ET.
\item The 2026 NFL Draft will be broadcast on ABC.
\item The 2026 NFL Draft will be broadcast on NFL Network.
\item The 2026 NFL Draft will be broadcast on ESPN Deportes.
\item Streaming options for the 2026 NFL Draft are available on NFL+.
\item Streaming options for the 2026 NFL Draft are available on the ESPN App.
\end{enumerate}

\begin{figure*}[!t]
\begin{lstlisting}[caption={Prompt used for claim extraction from AIOs},label={lst:claim-extraction-prompt},basicstyle=\ttfamily\scriptsize,breaklines=true,frame=single]
##### SYSTEM PROMPT

You are an assistant for fact-checkers.

You will be given the full text of a Google AI Overview. Your job is to extract all specific, verifiable, and fully decontextualized factual claims from it.

A claim must satisfy ALL of the following:
1. Verifiable: it can in principle be checked true or false against evidence.
2. Specific: it states a concrete fact, event, attribute, relationship, quantity, date, ranking, or action.
3. Decontextualized: it is fully understandable on its own, and its meaning in isolation matches its meaning in the AI Overview.
4. Entailed: if the AI Overview is true, the claim must also be true.

Rules:
1. Extract only claims that are explicitly supported by the AI Overview text. Do not use outside knowledge.
2. Do not invent or normalize missing details. If the text is vague, keep the claim equally vague or omit it.
3. If a statement is generic, normative, speculative, promotional, advisory, subjective, or otherwise not specifically verifiable, do not extract it.
4. If a sentence contains both generic language and one buried specific fact, extract only the specific fact.
5. If the text says that a person, organization, government body, report, court, source, or expert said, reported, announced, recommended, warned, found, highlighted, or did something, preserve that attribution when it is part of the meaning.
6. Resolve references when the text clearly supports it:
   - replace pronouns or shorthand with the fully specified referent when recoverable from nearby context;
   - expand partial names only when the full name is present in the AI Overview;
   - otherwise leave them unresolved only if the claim is still understandable and faithful.
7. If a statement has multiple plausible interpretations and the AI Overview does not clearly resolve the ambiguity, do not extract a claim from that ambiguous part.
8. Split multi-fact sentences into the simplest discrete factual claims that remain natural and useful for fact-checking.
9. Do not extract duplicate claims or near-duplicates.
10. Do not include citations, source names, bullet labels, headings, or formatting artifacts unless they are themselves part of a factual claim.

What to omit:
- opinions, praise, hype, or value judgments
- advice, instructions, recommendations to the reader
- vague trend language without a checkable proposition
- rhetorical summaries
- section headers like "Key takeaways" or "Why it matters"
- claims whose factual meaning depends on unresolved ambiguity

Examples of bad outputs:
- "This is a major development."
- "The product is impressive."
- "Experts think this is important."
- "The company may benefit in the future."

Examples of good behavior:
- If the text says "John Smith said the law would take effect in July 2026," extract:
  "John Smith said the law would take effect in July 2026."
  Do NOT extract:
  "The law would take effect in July 2026."
  unless the overview clearly presents that timing as a fact independent of John Smith's statement.

- If the text says "The update adds live translation and battery improvements," extract two claims:
  "The update adds live translation."
  "The update adds battery improvements."

- If the text says "The council expects the vote to happen next month," and the month is not recoverable from the overview, do NOT rewrite it with a calendar month.

Return ONLY a valid JSON object matching this shape:
{"claims":["claim 1","claim 2"],"no_claim_reason":""}

If there are no extractable claims, return:
{"claims":[],"no_claim_reason":"brief explanation of why the text does not contain extractable factual claims"}

If you output one or more claims, set "no_claim_reason" to an empty string.



##### USER PROMPT
AI Overview text:
{aio_text}

\end{lstlisting}
\end{figure*}

\section{Verification Prompt}
\label{app:verify_claims_prompt}

Listing~\ref{lst:claim-verification-prompt} presents the prompt used for verifying claims in AIOs. 

\begin{figure*}[!t]
\begin{lstlisting}[caption={Prompt used for claim extraction verification in AIOs},label={lst:claim-verification-prompt},basicstyle=\ttfamily\scriptsize,breaklines=true,frame=single]
Objective:
You are a content representativeness checker. You analyze whether an AI Overview (AIO) faithfully represents the information from its cited reference sources.

You will be given a list of factual claims extracted from an AI Overview, along with the text content of all reference sources cited by that AI Overview. For each claim, determine which reference(s) are relevant, whether they support or contradict the claim, and assign one of the following labels:

  CLEAR: The claim is clearly and directly supported by the reference content.
  VAGUE: The claim is supported but the reference content describes it in broader or less precise terms.
  AMBIGUOUS: The reference sources contain contradictory information about the claim. For example, one source supports the claim while another contradicts it.
  INCORRECT: The reference content directly contradicts the claim.
  OMITTED: The reference content does not mention the claim at all.

Here are some examples of individual claim assessments:

Claim: "The Tesla Model 3 starts at $38,990 in the United States."
Reference: "The Model 3 rear-wheel drive variant has a base price of $38,990."
-> CLEAR -- reference directly states the exact price.

Claim: "The iPhone 16 has an improved camera system."
Reference: "Apple's latest smartphone features several hardware upgrades."
-> VAGUE -- reference mentions upgrades but not the camera specifically.

Claim: "The event will be held in New Orleans."
Reference 1: "...takes place at the Caesars Superdome in New Orleans."
Reference 2: "...scheduled for MetLife Stadium in New Jersey."
-> AMBIGUOUS -- sources contradict each other.

Claim: "Python 4.0 was released in January 2026."
Reference: "There are currently no plans for a Python 4.0 release."
-> INCORRECT -- reference directly contradicts the claim.

Claim: "The restaurant offers free parking."
Reference: "Open Monday-Saturday, serving Italian and Mediterranean cuisine."
-> OMITTED -- reference does not mention parking at all.

Now verify ALL of the following claims against the reference sources below.

Keyword: "{keyword}"

Claims:
{claims_section}
Cited text snippets from the AI Overview (for additional context):
{cited_section}
Reference Sources:

{refs_section}
Output ONLY a valid JSON array. Each element must have the following fields:
[
  {
    "claim_id": 1,
    "claim": "<the claim text>",
    "label": "<CLEAR | VAGUE | AMBIGUOUS | INCORRECT | OMITTED>",
    "confidence": <0.0-1.0, how confident you are in this label>,
    "matched_references": ["R1", "R2"],
    "evidence": "<key phrase(s) from the reference that support your judgment>",
    "reasoning": "<brief explanation>"
  },
  ...
]

Rules:
1. You MUST output exactly one entry per claim, in the same order as the claims above.
2. matched_references should list ALL reference IDs (R1, R2, etc.) that are relevant to the claim. Use an empty list [] for OMITTED.
3. evidence should quote or closely paraphrase the specific text from references. Use "No relevant content found" for OMITTED.
4. confidence reflects how clearly the references support your judgment (1.0 = unambiguous match/contradiction, 0.5 = borderline).
5. CRITICAL: When a claim contains numbers, dates, or attributes paired with specific entities, verify that each value is assigned to the CORRECT entity. If the claim assigns value A to entity X and value B to entity Y, but the reference assigns value A to entity Y and value B to entity X, that is INCORRECT -- the values are swapped. Do not label a claim CLEAR just because the same numbers appear in the reference; check that they are attributed to the same things.
6. Only answer with the specified JSON array, no other text.
\end{lstlisting}

\end{figure*}
\section{Repeated Queries}
\label{app:repeated_queries}
\paragraph{Identification of repeated queries.}Each daily query set represents an independent snapshot of queries trending during the preceding 24 hours. Queries were not deliberately repeated within a daily snapshot. However, a query was issued again if it reappeared in a later Google Trends snapshot, and each execution was retained as a separate day-level observation because AIOs may change over time. We normalized query strings by applying Unicode normalization, converting them to lowercase, trimming whitespace, and collapsing repeated whitespace.

We analyzed 55,393 query executions, each representing one recorded submission to Google Search, covering 32,608 unique normalized queries. We divided these queries into four mutually exclusive groups: 21,605 (66.26\%) were executed only once; 3,038 (9.32\%) were repeated only within the same daily snapshot, primarily because they appeared under multiple topical categories (e.g., both ``Sports'' and ``Entertainment''); 6,847 (21.00\%) appeared on multiple dates but were executed at most once per date; and 1,118 (3.43\%) exhibited both cross-day and same-day repetition. Thus, 7,965 queries (24.43\%) appeared on at least two different dates, accounting for 27,633 executions (49.89\% of all executions). Same-day repetition produced 4,918 executions beyond the first execution of each date--query pair, representing 8.88\% of the dataset.

\paragraph{Query-balanced robustness analysis.}
To prevent frequently recurring queries from receiving disproportionate influence, we re-estimated the main results using query-balanced weights. For a query \(q\) with \(m_q\) eligible observations, each observation received weight \(w_{qi}=1/m_q\), so that every unique query had a total observation weight of one. For the AIO activation rate, \(m_q\) includes all executions of the query. For metrics describing AIO content or sources, \(m_q\) includes only observations in which an AIO was triggered. Claims, references, and source URLs inherit the weight of their parent AIO observation. Thus, query balancing removes the influence of query recurrence while retaining variation across repeated AIO observations.

\begin{table}[t]
  \centering
  \small
  \begin{tabular}{l r r r}
    \toprule
    Metric                   & Baseline & Balanced & Difference \\
    \midrule
    AIO activation (\%)      & 13.69  & 16.03  & +2.34 pp \\
    References per AIO       &  8.07  &  7.96  & $-0.11$ \\
    AIO-source UGC share (\%)& 14.18  & 14.35  & +0.17 pp \\
    AIO-source PC1           & 0.7319 & 0.7299 & $-0.0021$ \\
    References with ads (\%) & 50.63  & 52.39  & +1.75 pp \\
    Fidelity (\%)            & 88.97  & 89.11  & +0.14 pp \\
    Incorrect claims (\%)    &  2.66  &  2.63  & $-0.03$ pp \\
    Omitted claims (\%)      &  6.98  &  6.78  & $-0.20$ pp \\
    Top-20 domain share (\%) & 37.80  & 37.83  & +0.03 pp \\
    \bottomrule
  \end{tabular}
  \caption{Robustness of AIO-related results to query-balanced
    weighting.}
  \label{tab:query-balanced-aio}
\end{table}

As shown in Table~\ref{tab:query-balanced-aio}, query balancing produces only small changes in the AIO content, source-quality, fidelity, and source-concentration estimates. The largest difference is a 2.34-percentage-point increase in the estimated AIO activation rate; the other percentage-based estimates change by no more than 1.75 percentage points. The substantive conclusions therefore do not depend on the greater representation of recurring queries.

\paragraph{Variation across repeated observations.}We additionally compared chronologically adjacent observations of the same query. Across 17,867 cross-day adjacent pairs, the AIO-trigger outcome agreed in 95.92\% of comparisons. However, conditional on both observations triggering an AIO, only 0.06\% contained exactly the same normalized AIO text and only 0.24\% contained exactly the same set of cited URLs. The mean URL-set Jaccard similarity was 0.157. Thus, recurring queries generally did not mechanically duplicate the same AIO response or source set; instead, they captured temporal variation in AIO generation. Together with the query-balanced robustness results, these comparisons indicate that repeated queries do not drive our main findings.
\section{Webpage Changes}
\label{app:realtime_queries}
A subset of queries trigger AIOs that report a continuously-updating value, such as current weather, school-closure status, live product availability, app-status checks. 
As our crawler captures the cited pages at a different point in time than the AIO synthesis, the page text may shift by the time we read it, and the verifier may erroneously label the claim \texttt{Omitted} or \texttt{Incorrect}.
We identify such queries via a heuristic. 
Specifically, we first manually review a sample of failed verifications and extract seed keywords, and then automatically identify claims that use such keywords. 
For example, for the Climate category (where the pattern is most concentrated), we identified keywords such as \texttt{weather} and \texttt{forecast}.

Resulting exclusions included 12 Jobs \& Education AIOs comprising 107 claims, mainly associated with school-closure and locality queries such as \texttt{school closings for tomorrow}; 12 Technology AIOs comprising 126 claims, associated with live weather, app-status, and imminent-release queries such as \texttt{google weather} and \texttt{tomodachi life release time}; and 1 Shopping AIO comprising 15 claims for \texttt{costco finds}.
Note that as we rely on a keyword-based heuristic, we are prone to making mistakes, including over- and under-estimating claims pertaining to real-time data on webpages.

\end{document}